# Discussions on non-probabilistic convex modelling for uncertain problems

B.Y. Ni [1], C. Jiang [1,*], Z.L. Huang [2]

[1] *State Key Laboratory of Advanced Design and Manufacturing for Vehicle Body, College of Mechanical and Vehicle Engineering, Hunan University, Changsha, P. R. China 410082*

[2] *School of Mechanical and Electrical Engineering, Hunan City University, YiYang City, P. R. China 413002*

**Abstract**

Non-probabilistic convex model utilizes a convex set to quantify the uncertainty domain of uncertain-but-bounded parameters, which is very effective for structural uncertainty analysis with limited or poor-quality experimental data. To overcome the complexity and diversity of the formulations of current convex models, in this paper, a unified framework for construction of the non-probabilistic convex models is proposed. By introducing the correlation analysis technique, the mathematical expression of a convex model can be conveniently formulated once the correlation matrix of the uncertain parameters is created. More importantly, from the theoretic analysis level, an evaluation criterion for convex modelling methods is proposed, which can be regarded as a test standard for validity verification of subsequent newly proposed convex modelling methods. And from the practical application level, two model assessment indexes are proposed, by which the adaptabilities of different convex models to a specific uncertain problem with given experimental samples can be estimated. Four numerical examples are investigated to demonstrate the effectiveness of the present study.

# 1 Introduction

The physical parameters used to describe a structure or system in practical engineering are often uncertain due to geometrical imperfections, model inaccuracies, external interferences, etc. As the principal way for quantification of these physical uncertainties, the probabilistic methods have been successfully applied to various engineering problems. However, in practical engineering, the credible probability models may not available if experimental data are insufficient [1]. To deal with the difficulty that sufficient information on the uncertain parameters are often unavailable due to limitations of test conditions or cost in practical engineering, the non-probabilistic convex modelling approach [2-6], has been developed since the early 1990s. The non-probabilistic convex modelling approach advocates representation of uncertain parameters by a convex set, which relies upon knowledge of the variation bounds of the parameters. Compared to the precise probability distribution functions, the variation bounds are much easier to obtain, since it generally needs only a small number of experimental data or just the experience of engineers.

It should be pointed out that convex model approach does not represent only a single model;

instead it represents a model set. Presently, interval model and ellipsoid model are two kinds of the most commonly used models in this area. In interval model, the uncertainty of a single variable is described through its upper and lower bounds. The uncertainty domain is then depicted as a 'multidimensional box' for a multidimensional problem. For ellipsoid model, it is assumed that the parametric uncertainty lies within a 'multidimensional ellipsoid'. The degree of uncertainty and the degree of correlation of the variables are described by the size and shape of the ellipsoid. In theory, interval model can deal only with problems involving independent uncertain variables, while the ellipsoid model can deal only with dependent variables. Based on interval model, the anti-optimization analysis [7-10] was performed to seek for the least favorable response of a structural system under imposed constraints of uncertain-but-bounded parameters. Under the assumption of small uncertainty, a first-order interval perturbation method was applied to determine the influence of interval parameters on eigenvalues [11] and static displacements [12] of structures. By taking into account the actual variation and dependency of uncertain parameters affecting the mass and stiffness matrices, the lower and upper bounds of the natural frequencies of a structure with uncertain-but-bounded parameters were evaluated [13]. To provide a measure for the individual influence of interval inputs on the range of the obtained interval outcome of structural systems, the interval sensitivity analysis was studied and applied to the envelope frequency response function analysis of uncertain mechanical structures [14]. To find the best ellipsoidal convex model fitting the given experimental data of uncertain parameters, a Gram-Schmidt orthogonalization procedure for rotation of the coordinate system was put forwarded [15]. To improve the efficiency in constructing the multidimensional ellipsoid, a correlation analysis technique was proposed [16], providing a fundamental mathematical tool for convex modelling. For the same purpose, an equivalent semi-definite programming (SDP) formulation was given to determine the minimum-volume ellipsoid model, based on which a reliability-based topology optimization method was further developed [17]. A non-probabilistic reliability model was created for structures with convex model uncertainty, and two efficient computational methods, namely the first order approximation method (FOAM) and the second order approximation method (SOAM) were also formulated for this reliability model [18]. As two typical convex models for uncertainty quantification, the ellipsoid model and the interval model were compared for carrying out dynamic response analysis and buckling failure analysis of bars [19]. Recently, some variants of the traditional convex models were also developed, including the multi-ellipsoid model [20, 21], the multidimensional parallelepiped model [22, 23], the super-ellipsoidal model [24], the convex model process [25], etc. Also, the convex model approach presently has been applied in various areas in structural mechanics, such as finite element analysis [26, 27], non-probabilistic reliability analysis [5, 18, 28-31], uncertain optimization [32-37], hybrid reliability analysis [38-41], etc.

Though as mentioned above some convex modelling methods have been successfully developed, overall research in this area is still very weak. So far, a systematical research on mathematical principles or fundamentals is still absent for uncertainty quantification of convex model approach. For example, the lack of evaluation standards on convex modelling leads researchers to create models by following their own rules, in many cases causing the uncertainty analysis process complicated and confusing. From the view of engineers, the lack of mathematical evaluation criteria also makes it unclear that which kind of model is more appropriate to a specific problem. Furthermore, various convex models presently have been proposed in this area, but these models are given in quite different forms, to a large extent bringing about inconvenience to the application of convex model approach in practical engineering.

For the above reasons, a unified construction framework for convex models is proposed by introducing the correlation analysis technique; more importantly, an evaluation criterion for convex modelling methods is also suggested, which can be regarded as a test standard for subsequent newly developed models in this area. Specifically, the construction procedure of the multidimensional ellipsoid (ME) model and the multidimensional parallelepiped (MP) models are unified. In this way, the only differences between different models are the parametric correlation quantifications and the formulation of the *Characteristic Matrices* in their analytical expressions defined in this paper. Subsequently, the mathematical property named as "unbiasedness" is proposed as a primary evaluation criterion for convex modelling methods, by which theoretic validity verification can be performed for different convex models. The reminders of this paper are then organized as follows: in Section 2, two kinds of convex models considering correlation, namely the ME model and the MP models are discussed and a unified construction method is proposed. In Section 3, the evaluation criteria for verification of the convex modelling methods and the model assessment indexes for practical applications are proposed. Numerical examples are investigated in Section 4 before we draw conclusions in Section 5.

## 2 Two kinds of convex models considering correlation

Assume that there exist $n$ uncertain parameters $X_i$, $i=1,2,\cdots,n$ with $N_s$ groups of experimental samples $\mathbf{x}^{(s)}$, $s=1,2,\cdots,N_s$, which, for a practical structure, could be material properties, external loads, geometrical sizes, etc. These parameters constitute an $n$-dimensional parameter space, namely $X$-space. By using the convex model approach, the variation of each uncertain parameter $X_i$, $i=1,2,\cdots,n$ is quantified by an interval, namely

$$X_i \in X_i^I = [X_i^L, X_i^U] = [X_i^m - X_i^r, X_i^m + X_i^r], i=1,2,\cdots,n \tag{1}$$

where the superscripts $I$, $L$ and $U$ denote interval, lower bound and upper bound, respectively; $X_i^m$ and $X_i^r$ are midpoint and radius, respectively. This kind of uncertain parameters can also be called as interval variables.

The uncertainty domain of multiple interval variables $\mathbf{X}=[X_1, X_2, \cdots, X_n]^{\mathrm{T}}$ constitutes a bounded set, denoted as $\Omega_X$ in this paper. Based on the uncertainty domain $\Omega_X$, the *marginal interval* of $X_i$ is defined as

$$X_i^I = [X_i^L, X_i^U] = \{X_i | \mathbf{X} \in \Omega_X\}, i=1,2,\cdots,n \tag{2}$$

which indicates the variation range of $X_i$ regardless of the values of other interval variables.

Specifically, the marginal interval $X_i^I = [-1,1]$ is also called as a *standard marginal interval*; and the convex set with standard marginal intervals is called as a *regularized convex set*. Generally, in convex modelling approach, the primary task is to construct the uncertainty domain for interval variables with given marginal intervals.

If the interval variables are independent, the uncertainty domain of $\mathbf{X} = [X_1, X_2, \cdots, X_n]^T$, constitutes a hyper-rectangle $\bar{\Omega} = [X_1^L, X_1^U] \times [X_2^L, X_2^U] \times \cdots \times [X_n^L, X_n^U]$ in the $X$-space. If correlation exists among the parameters, generally a convex subset of $\bar{\Omega}$ can be used to describe the uncertainty domain, such as the multidimensional ellipsoid (ME) model [2, 3], the multidimensional parallelepiped (MP) models [22, 23], etc. Given the marginal intervals $X_i^I$, $i = 1, 2, \cdots, n$ and the experimental samples $\mathbf{x}^{(s)}$, $s = 1, 2, \cdots, N_s$, the ME model, as shown in Fig. 1(a), employs an *n*-dimensional ellipsoid enveloping the samples to represent the uncertainty domain

$$\Omega_X = \left\{ \mathbf{X} \big| (\mathbf{X} - \mathbf{X}^m)^T \mathbf{G}_E (\mathbf{X} - \mathbf{X}^m) \leq 1 \right\} \tag{3}$$

where $\mathbf{X}^m = [X_1^m, X_2^m, \cdots, X_n^m]^T$ denotes the vector of midpoints, and the *Characteristic Matrix* $\mathbf{G}_E$ of the ME model is an invertible, symmetric positive-definite matrix, which determines the size and orientation of the multidimensional ellipsoid. As shown in Fig. 1(b), the MP model suggests an *n*-dimensional parallelepiped for description of the uncertainty domain, which can be expressed as

$$\Omega_X = \left\{ \mathbf{X} \big| \big[ |\mathbf{G}_P (\mathbf{X} - \mathbf{X}^m)| \big] \leq \mathbf{e} \right\} \tag{4}$$

or

$$\Omega_X = \left\{ \mathbf{X} \big| -\mathbf{e} \leq \mathbf{G}_P (\mathbf{X} - \mathbf{X}^m) \leq \mathbf{e} \right\} \tag{5}$$

where the *Characteristic Matrix* $\mathbf{G}_P$ of the MP model determines the shape and size of the multidimensional parallelepiped, and $\mathbf{e} = [1 \ 1 \ \cdots \ 1]^T$ is an *n*-dimensional vector whose elements are all 1. From Eqs. (3) and (4) it can be seen that once the *Characteristic Matrix* $\mathbf{G}_E$ of the ME model or $\mathbf{G}_P$ of the MP model is created, the uncertainty domain $\Omega_X$ of the uncertain parameters $X_i$, $i = 1, 2, \cdots, n$ can be determined.

## 2.1 Construction of ME model and MP model

In this section, a unified procedure for constructing the ME and the MP models is proposed. A correlation analysis technique [16] is introduced here, by which the enormous complexity existing in the construction of multidimensional convex models can be greatly alleviated. Firstly, the correlation quantification of interval variables in the ME model and the MP models are introduced, and then the analytical expressions of both models are given. Based on this, a unified construction procedure for non-probabilistic convex models is finally derived.

### 2.1.1 Regularization of interval variables

Given the marginal intervals $X_i^I$ and the experimental samples $\mathbf{x}^{(s)}$, $s = 1, 2, \cdots, N_s$, the first step in our proposed construction procedure is regularization. An interval variable with $X^m = 0$ and $X^r = 1$ is referred to as *standard interval variable*. The transformation from an interval variable $X_i$ to a standard interval variable $U_i \in U_i^I = [-1, 1]$ is named as *regularization*

$$U_i = \frac{X_i - X_i^m}{X_i^r} \tag{6}$$

For an *n*-dimensional problem the following matrix form can be used

$$\mathbf{U} = \mathbf{D}_X^{-1}(\mathbf{X} - \mathbf{X}^m) \tag{7}$$

where $\mathbf{D}_X = \mathrm{diag}\{X_1^r, X_2^r, \ldots, X_n^r\}$. Correspondingly, the samples $\mathbf{x}^{(s)}$, $s = 1, 2, \cdots, N_s$ of the interval variables can be transformed in the same way

$$\mathbf{u}^{(s)} = \mathbf{D}_X^{-1}(\mathbf{x}^{(s)} - \mathbf{X}^m) \tag{8}$$

By the above regularization, the problem of constructing a multidimensional convex set $\Omega_X$ with given marginal intervals $X_i^I, i = 1, 2, \cdots, n$ in *X*-space is equivalent to finding a convex set $\Omega_U$ with standard marginal intervals in *U*-space. The mathematical description of $\Omega_X$ can be easily derived from $\Omega_U$ by just an inverse transformation of that in Eq. (7).

### 2.1.2 Multidimensional ellipsoid (ME) model

The ME model depicts the uncertainty domain of two interval variables by an ellipse and that of *n* variables by an *n*-dimensional ellipsoid, as given in Eq. (3). Traditionally, the best ME model is identified as the one enveloping all the given experimental samples but has a minimum volume, which theoretically can be found by solving an optimization problem [16]. However, actually this "minimum volume method" can work well only for some simple problems with a small number of variables and less sample size. To overcome this difficulty, a correlation analysis technique was proposed for ME modelling in the authors' previous work [16]. Based on the correlation analysis technique, the ellipsoidal domain $\Omega_X$ of a general *n*-dimensional problem can be efficiently created through transforming the *n*-dimensional problem into a set of bivariate sub-problems that are easy to deal with. Inspired by this correlation analysis technique, we will develop a unified convex modelling framework. Firstly, a concept for non-probabilistic correlation quantification of two interval variables is defined as follows.

**Definition 1.** For standard interval variables $U_i$ and $U_j$ with samples $\{(u_i^{(s)}, u_j^{(s)})\}, s = 1, 2, \ldots, N_s$, the correlation coefficient that determined by the geometric characteristics of a regularized convex set enclosing all the samples is named as *Convex Correlation Coefficient* or CCC in brief, denoted as $r_c(U_i, U_j)$.

In the cases without confusion, the CCC $r_c(U_i, U_j)$ can be also denoted as $r_{c\text{-}ij}$ or $r_c$ in brief. Theoretically, the value of CCC should satisfy $-1 \leq r_c \leq 1$. A positive value of $r_c$ indicates positive correlation between $U_i$ and $U_j$, a negative value indicates negative correlation, and zero means uncorrelated. The stronger is the correlation between the pair of interval variables, the larger will be the absolute value of $r_c$. For an ME model, the definition of CCC is given in the following way.

**Definition 2.** For standard interval variables $U_i$ and $U_j$ with samples $\{(u_i^{(s)}, u_j^{(s)})\}$, $s = 1, 2, \ldots, N_s$, assume that the minimum-area ellipse with standard marginal intervals enclosing all the samples exists, then the Convex Correlation Coefficient $r_c$ of $U_i$ and $U_j$ in ME model is defined as

$$r_c = \frac{b^2 - a^2}{b^2 + a^2} \tag{9}$$

where $a$ denotes the semi-axis length of this ellipse in the axis of $U_i = -U_j$, and $b$ denotes the semi-axis length in the axis of $U_i = +U_j$, as shown in Fig. 2(a), in which the small circles represent the samples.

Figure 2(a), (b) and (c) presents the shapes of the ellipse when the CCC is positive, zero and negative. Herein as an example, the figures exactly depict the cases of $r_c = +0.6, 0, -0.6$, respectively.

**Lemma 1.** The regularized ellipse-shaped uncertainty domain as shown in Fig. 2 can be analytically expressed as

$$\begin{bmatrix} U_i & U_j \end{bmatrix} \begin{bmatrix} 1 & r_c \\ r_c & 1 \end{bmatrix}^{-1} \begin{bmatrix} U_i \\ U_j \end{bmatrix} \leq 1 \tag{10}$$

where $-1 < r_c < 1$ is defined in Definition 2.

For an $n$-dimensional problem with standard interval variables $\mathbf{U} = [U_1, U_2, \cdots, U_n]^T$, after the CCCs between each pair of interval variables are obtained, a symmetrical correlation matrix $\mathbf{R}$ then can be constituted

$$\mathbf{R} = \begin{bmatrix} r_{c\text{-}11} & r_{c\text{-}12} & \cdots & r_{c\text{-}1n} \\ r_{c\text{-}21} & r_{c\text{-}22} & \cdots & r_{c\text{-}2n} \\ \vdots & \vdots & \vdots & \vdots \\ r_{c\text{-}n1} & r_{c\text{-}n2} & \cdots & r_{c\text{-}nn} \end{bmatrix} \tag{11}$$

in which $r_{c\text{-}ij}$ denotes the CCC between interval variables $U_i$ and $U_j$. Then the $n$-dimensional ellipsoid-shaped uncertainty domain $\Omega_U$ can be analytically expressed as

$$\Omega_U = \{\mathbf{U} | \mathbf{U}^T \mathbf{R}^{-1} \mathbf{U} \leq 1\} \tag{12}$$

**Lemma 2.** The projection of the $n$-dimensional uncertainty domain $\Omega_U$ given in Eq. (12) on the coordinate plane of $U_i - U_j$ is an ellipse which can be expressed as

$$\begin{bmatrix} U_i & U_j \end{bmatrix} \begin{bmatrix} 1 & r_{c-ij} \\ r_{c-ij} & 1 \end{bmatrix}^{-1} \begin{bmatrix} U_i \\ U_j \end{bmatrix} \leq 1 \tag{13}$$

The proof is given in Appendix A.

From Lemma 2 it can be concluded that the $n$-dimensional ellipsoid-shaped uncertainty domain $\Omega_U$ can be rebuilt from its 2-dimensional information, namely, its projections on the 2-dimensional coordinate planes. Hence to construct a multi-dimensional ellipsoid from the 2-dimensional spaces to the $n$-dimensional space is feasible. Combining with Lemma 1, it can also be concluded that the marginal intervals of the $n$-dimensional ellipsoid-shaped uncertainty domain $\Omega_U$ are exactly accord to $U_i^I = [-1, 1]$, $i = 1, 2, \cdots, n$.

With the created $n$-dimensional uncertainty domain $\Omega_U$ for standard interval variables $\mathbf{U}=[U_1,U_2,\cdots,U_n]^T$, the uncertainty domain $\Omega_X$ for the original interval variables $\mathbf{X}=[X_1,X_2,\cdots,X_n]^T$ can be then derived only by substituting Eq. (7) into Eq. (12), which yields

$$\Omega_X = \left\{\mathbf{X} \big| (\mathbf{X}-\mathbf{X}^m)^T \mathbf{C}_X^{-1}(\mathbf{X}-\mathbf{X}^m) \leq 1\right\} \tag{14}$$

in which $\mathbf{C}_X$ is defined as the covariance matrix

$$\mathbf{C}_X = \mathbf{D}_X \mathbf{R} \mathbf{D}_X \tag{15}$$

From the above analyses, it is not difficult to find that the *Characteristic Matrix* $\mathbf{G}_E$ of the multidimensional ellipsoid can be determined through $\mathbf{G}_E = \mathbf{C}_X^{-1}$.

### 2.1.3 Multidimensional parallelepiped (MP) model

Different from the ME model, the MP model utilizes a parallelogram to quantify the uncertainty of two interval variables and a multidimensional parallelepiped for multiple intervals, as shown in Fig. 1(b) and analytically expressed by Eq. (4). The MP model was initially proposed to overcome the deficiency of the ME model in uncertainty quantification of complex multi-source uncertainty problems [22], in which an affine coordinate system was employed to describe the uncertainty domain. It was improved in Ref. [23] and the analytical expression for MP model was successfully created. In this section, besides the improved model in Ref. [23], several new MP models by using different correlation quantification methods are also proposed. Although different in quantifications of correlation and uncertainty, these MP models share the same mathematical expression, namely, Eq. (4), and can be constructed in a similar procedure. For this reason, in the following text we will give detailed descriptions to one of the MP models and only brief introductions to the others.

*MP-II model*

The MP-II model is a further improved model basing on that proposed in Ref. [23]. The correlation between the two standard interval variables $U_i$ and $U_j$ is also quantified by a rhomb enclosing all the samples, but the CCC $r_c$ is redefined.

**Definition 3.** For standard interval variables $U_i$ and $U_j$ with samples $\{(u_i^{(s)}, u_j^{(s)})\}$, $s=1,2,\ldots,N_s$, assume that the minimum-area rhomb with standard marginal intervals enclosing all the samples exists, then the Convex Correlation Coefficient $r_c$ of $U_i$ and $U_j$ in MP-II model is defined as the same formula as Eq. (9), while the semi-axis lengths $a$ and $b$ are as shown in Fig. 3(a).

The stronger is the correlation between interval variables $U_i$ and $U_j$, the more compact will be the rhomb. Figure 3 shows the shapes of three 2-dimensional cases of the uncertainty domain depicted by MP-II model; herein the CCCs are $r_c = +0.6, 0, -0.6$, respectively.

For an $n$-dimensional problem with standard interval variables $\mathbf{U}=[U_1,U_2,\cdots,U_n]^T$, the correlation matrix $\mathbf{R}$ can be constituted by Eq. (11) when the CCCs between each pair of interval variables are obtained through the way of Definition 3. With the correlation matrix $\mathbf{R}$, the analytical expression of the MP-shaped uncertainty domain $\Omega_U$ can be then formulated with the following definition.

**Definition 4.** The unified formulation of the MP-shaped uncertainty domain $\Omega_U$ for standard interval variables $\mathbf{U} = [U_1, U_2, \cdots, U_n]^T$ is defined as

$$\Omega_U = \left\{ \mathbf{U} \middle| \left[\left|\mathbf{S}^{-1}\mathbf{U}\right|\right] \leq \mathbf{e} \right\} \tag{16}$$

where $[|\cdot|]$ constitutes a vector or matrix composed by taking absolute value to each element of the vector or matrix inside. The $n \times n$ matrix $\mathbf{S}$ is referred to as *Shape Matrix* and defined as

$$\begin{aligned} \mathbf{S} &= \mathbf{TH} \\ \mathbf{T} &= \text{diag}\begin{bmatrix} w_1 & w_2 & \cdots & w_n \end{bmatrix} \\ w_i &= \frac{1}{\sum_{j=1}^n |\mathbf{H}(i,j)|}, i = 1, 2, \cdots, n \\ \mathbf{e} &= \begin{bmatrix} 1 & 1 & \cdots & 1 \end{bmatrix}_{n \times 1}^T \end{aligned} \tag{17}$$

The $n \times n$ matrix $\mathbf{H}$ is defined as the *Core of Shape Matrix* (CSM), which can be derived from the correlation matrix $\mathbf{R}$. For the MP-II model, the CSM $\mathbf{H}$ is defined as the square root of the correlation matrix $\mathbf{R}$, namely

$$\mathbf{H} = \mathbf{R}^{1/2} \tag{18}$$

Here in the MP-II model, the CSM $\mathbf{H}$ is symmetrical and its computational method and algorithm can be referred to Ref. [42]. Specifically, when only two standard interval variables $U_i$ and $U_j$ are involved, the two-dimensional uncertainty domain depicted in Fig. 3 can be expressed as

$$\Omega_U = \left\{ \begin{bmatrix} U_i \\ U_j \end{bmatrix} \middle| \left[\left|\mathbf{S}^{-1}\begin{bmatrix} U_i \\ U_j \end{bmatrix}\right|\right] \leq \begin{bmatrix} 1 \\ 1 \end{bmatrix} \right\} \tag{19}$$

which can be conveniently verified by its geometric characteristics.

**Lemma 3.** The marginal intervals of the uncertainty domain $\Omega_U$ depicted by the MP model are exactly equal to $U_i^I = [-1, 1]$, $i = 1, 2, \cdots, n$.

Lemma 3 indicates that a multidimensional parallelepiped-shaped uncertainty domain constructed by Eq. (16) will always be valid, in view of the marginal intervals. With the created uncertainty domain $\Omega_U$ for standard interval variables $\mathbf{U} = [U_1, U_2, \cdots, U_n]^T$, the uncertainty domain for the original interval variables $\mathbf{X} = [X_1, X_2, \cdots, X_n]^T$ can be then derived by substituting Eq. (7) into Eq. (16), which yields

$$\Omega_X = \left\{ \mathbf{X} \middle| \left[\left|(\mathbf{D}_X \mathbf{S})^{-1}(\mathbf{X} - \mathbf{X}^m)\right|\right] \leq \mathbf{e} \right\} \tag{20}$$

Thus, the *Characteristic Matrix* $\mathbf{G}_P$ in Eq. (4) of the MP-II model can be determined as $\mathbf{G}_P = (\mathbf{D}_X \mathbf{S})^{-1}$.

As mentioned before, in this paper we will create and discuss several different MP models, and they depict the uncertainty domain by different shapes of multidimensional parallelepipeds. However, the construction procedures of these MP models are similar. The only differences exist in two aspects, i.e., the quantification of correlation and the derivation of CSM. In the following, only the definitions of the CCC $r_c$ and the CSM $\mathbf{H}$ in other four MP models are presented.

*MP-I model*

The MP-I model was originally proposed in Ref. [23]. Different from the MP-II model, the CCC $r_c$ in MP-I model is defined as

$$r_c = \frac{b-a}{b+a} \tag{21}$$

where $a$ denotes the semi-axis length in the axis $U_i = -U_j$ and $b$ denotes the semi-axis length in the axis $U_i = +U_j$, as shown in Fig. 4(a). The three 2-dimensional cases of the MP-I model presented in Fig. 4 correspond to those when the CCC $r_c$ equals to +0.6, 0 and −0.6, respectively. The CSM **H** used to construct the MP-I model is just the correlation matrix **R**, namely

$$\mathbf{H} = \mathbf{R} \tag{22}$$

Note that although the MP-I model also uses a rhomb for correlation quantification, it differs with the MP-II model in definition of the CCC, which can also be illustrated by comparing Fig. 4 with the cases of the MP-II model in Fig. 3.

*Rectangular MP model*

In the Rectangular MP model, the correlation between the standard interval variables $U_i$ and $U_j$ is quantified by a rectangle, as shown in Fig. 5(a). The CCC $r_c$ of $U_i$ and $U_j$ in Rectangular MP model is defined as the same formula as Eq. (9), while the semi-axis lengths $a$ and $b$ are as shown in Fig. 5(a).

For an *n*-dimensional problem with standard interval variables $\mathbf{U} = [U_1, U_2, \cdots, U_n]^\mathrm{T}$, the CSM **H** in Rectangular MP model can be derived from the eigen-decomposition of the correlation matrix **R**

$$\begin{aligned} \mathbf{R} &= \mathbf{Q}\mathbf{\Lambda}\mathbf{Q}^\mathrm{T} \\ \mathbf{H} &= \mathbf{Q}\mathbf{\Lambda}^{1/2} \end{aligned} \tag{23}$$

The Rectangular MP model is included among the MP models, because that though the CCC between two standard interval variables is defined by a rectangle, the uncertainty domain of a multidimensional problem depicted by this model is in fact a multidimensional parallelepiped-shaped convex set. It can be also seen that independent interval variables cannot be well depicted by the Rectangular MP model, as shown in Fig. 5(b). Therefore, unlike the other MP models, theoretically the Rectangular MP model can only deal with dependent interval variables.

*Lower Triangle MP model (LTri-MP model)*

In the LTri-MP model, the correlation between two standard interval variables $U_i$ and $U_j$ is quantified by a parallelogram as shown in Fig. 6(a). In order to unify it into the MP modelling framework in Eq. (16), here the CCC $r_c$ of $U_i$ and $U_j$ is defined as

$$r_c = \frac{|1-t|}{\sqrt{(1-t)^2 + t^2}} \mathrm{sgn}(\cos\theta) \tag{24}$$

where *t* denotes a half-length of the vertical side, $\theta$ denotes the included angle in the lower left of the parallelogram as shown in Fig. 6(a), and sgn(·) is the signum function. Figure 6(a), (b) and (c) presents the shapes of three 2-dimensional cases in LTri-MP model when the CCC is positive, zero and negative, respectively.

For an *n*-dimensional problem with standard interval variables $\mathbf{U} = [U_1, U_2, \cdots, U_n]^T$, the CSM $\mathbf{H}$ in the LTri-MP model can be derived from the Cholesky decomposition of the correlation matrix $\mathbf{R}$

$$\begin{aligned} \mathbf{R} &= \mathbf{L}\mathbf{L}^T \\ \mathbf{H} &= \mathbf{L} \end{aligned} \quad (25)$$

where $\mathbf{L}$ is a lower triangular matrix.

### *Upper Triangle MP model (UTri-MP model)*

The uncertainty domain of two interval variables $U_i$ and $U_j$ depicted by the UTri-MP model is shown in Fig. 7(a). In UTri-MP model, the CSM $\mathbf{H}$ is an upper triangular matrix obtained by

$$\begin{aligned} \mathbf{R} &= \mathbf{U}\mathbf{U}^T \\ \mathbf{H} &= \mathbf{U} \end{aligned} \quad (26)$$

Figure 7(a), (b) and (c) presents the shapes of three 2-dimensional cases in the UTri-MP model when the CCC $r_c$ is positive, zero and negative, respectively.

### 2.1.4 A unified construction procedure

From the above discussions we can find that different convex models can be employed for uncertainty quantification of the uncertain-but-bounded parameters. Based on the correlation analysis technique, a multidimensional uncertainty domain can be easily constructed only by correlation analyses for multiple bivariate problems; and the total number of the bivariate problems is $C_n^2 = n(n-1)/2$. With the correlation matrix created, the analytical expression of a convex set-shaped uncertainty domain can be then determined simultaneously. A unified procedure is created to construct an ME or an MP model through a set of samples, and its diagrammatic sketch is provided in Fig. 8. The construction procedure is summarized as follows:

***Step 1:*** Data preparation. Obtain the marginal intervals $X_i^I$ of the uncertain-but-bounded parameters $X_i, i = 1, 2, \cdots, n$ and their experimental samples $\mathbf{x}^{(s)}, s = 1, 2, \cdots, N_s$.

***Step 2:*** Regularization. Map the uncertain parameters and the experimental samples from *X*-space into *U*-space, where the marginal interval of each variable $U_i$ needs to be regularized to a standard interval $U_i^I = [-1, 1]$.

***Step 3:*** Model selection. Select the ME model or one of the MP models for describing the uncertainty domain of the parameters.

***Step 4:*** Correlation evaluation. According to the convex model selected in Step 3, evaluate the Convex Correlation Coefficients (CCCs) $r_c$ for each pair of interval variables $(U_i, U_j), i < j$.

***Step 5:*** Correlation matrix establishment. With all the CCCs from Step 4, constitute the correlation matrix $\mathbf{R}$.

***Step 6:*** Derivation of the Core of Shape Matrix (CSM). According to the type of the selected convex model, create the CSM $\mathbf{H}$ through the correlation matrix $\mathbf{R}$. If the Ellipsoid model is selected, skip this step and jump to Step 8.

***Step 7:*** Formulation of the Shape Matrix. With the CSM $\mathbf{H}$, determine the Shape Matrix $\mathbf{S}$ by Eq.

(17).

*Step 8:* Formulation of the Characteristic Matrix. Create the Characteristic Matrix $\mathbf{G}_E$ for ME model or $\mathbf{G}_P$ for MP model.

*Step 9:* Analytic expression of the uncertainty domain. Based on the Characteristic Matrix, finally obtain the analytic expression of the uncertainty domain $\Omega_X$ of the parameters $X_i, i=1,2,\cdots,n$.

Through the above procedure, a convex set depicting the parametric uncertainty can be obtained with the given experimental data. Although the present method cannot guarantee to produce an uncertainty domain with the minimum volume, it generally ensures to give a reasonable and useful uncertainty domain for a practical engineering problem. More importantly, through introducing the correlation analysis technique, the complex optimization problem existing in the traditional convex modelling methods is successfully avoided, and hence theoretically our construction procedure is applicable to high-dimensional problems. Also, in our framework, different convex models can be adopted for quantification of the parametric uncertainty, which provides a very flexible treatment in practical uncertainty analysis and reliability design.

## 2.2 An alternative to Convex Correlation Coefficient

It can be found that the correlation matrix plays a key role in the construction of a convex model. In the above the correlation matrix is constituted by the CCCs between interval variables. To evaluate the CCC $r_{c-ij}$ between interval variables $U_i$ and $U_j$, a 2D convex set, which can be an ellipse or a parallelogram enclosing all the bivariate samples with the minimum area, need to be obtained. To avoid finding this minimum convex set in the 2-dimensional parameter space, the Sample Correlation Coefficient (SCC) can also be used as an alternative of CCC and can furtherly simplify the uncertainty modelling process of convex models.

### 2.2.1 Definition of the Sample Correlation Coefficient (SCC)

**Definition 5.** For interval variables $X_i \in X_i^I = [X_i^L, X_i^U]$ and $X_j \in X_j^I = [X_j^L, X_j^U]$ with $N_s$ groups of experimental samples $\left(x_i^{(s)}, x_j^{(s)}\right), s=1,2,\ldots,N_s$, the *Sample Correlation Coefficient* (SCC) between $X_i$ and $X_j$ is defined as

$$r_s = \frac{\sum_{s=1}^{N_s}\left(x_i^{(s)} - X_i^m\right)\left(x_j^{(s)} - X_j^m\right)}{\sqrt{\sum_{s=1}^{N_s}\left(x_i^{(s)} - X_i^m\right)^2 \sum_{s=1}^{N_s}\left(x_j^{(s)} - X_j^m\right)^2}} \tag{27}$$

On the left hand the subscript "s" represents "Sample Correlation Coefficient", while the superscript "*s*" on the right hand represents the sequence number of the experimental samples.

**Lemma 4.** $-1 \leq r_s \leq 1$. There is $r_s = \pm 1$ when and only when $\frac{x_i^{(s)} - X_i^m}{x_j^{(s)} - X_j^m} = \pm c$, $s=1,2,\cdots,N_s$.

Here *c* stands for a positive constant.

Lemma 4 can be easily proved by the well-known Cauchy-Schwarz inequality [43].

It should be pointed out that SCC provides a unified form for correlation measure of interval variables, since it is directly derived from the samples and has no relation with the type of the convex model.

**2.2.2 Comparison of SCC and CCC in convex modelling**

*Relationship*

**Lemma 5.** There is $r_s = r_c$ when the samples $\left(x_i^{(s)}, x_j^{(s)}\right), s = 1, 2, \cdots, N_s$ of interval variables $X_i$ and $X_j$ are uniformly distributed within a 2-dimensional convex set depicted by the ME or MP models (except for the MP-I model).

Lemma 5 indicates that for a certain convex set $\Omega_{ij}$, such as an ellipse, if the samples are uniformly distributed within this set $\Omega_{ij}$, the SCC $r_s$ calculated by Eq. (27) will be equal to the CCC $r_c$ obtained by Eq. (9). Similar conclusion holds for the MP models, except for the MP-I model. The proofs are given in Appendix B1-B2. The proofs illustrate that when the samples $\mathbf{u}^{(s)}$, $s = 1, 2, \cdots, N_s$ of the interval variables $\mathbf{X} = [X_1, X_2, \cdots, X_n]^T$ are uniformly distributed within the $n$-dimensional ellipsoid- or parallelepiped-shaped domain depicted by the ME model or the MP-model (except for the MP-I model), the SCC $r_{c\text{-}ij}$ of $X_i$ and $X_j$ calculated by the bivariate samples $\left(x_i^{(s)}, x_j^{(s)}\right), s = 1, 2, \cdots, N_s$ will be accordant with the corresponding value of the correlation matrix $\mathbf{R}$ in the analytical expression of the convex model. By setting $n = 2$, the conclusion of Lemma 5 can be derived. Note that here the uniform distribution is a sufficient condition, rather than a necessary condition.

*Differences*

The main differences between SCC and CCC in convex modelling can be summarized into two points. Firstly, in most general cases where experimental samples are scattered uniformly, the SCC can approximate the CCC well, as shown in Fig. 9. For a certain group of general samples, the CCC and the SCC-based two dimensional ellipsoid models are presented in Fig. 9(a) and (b). By comparison it can be found that the ellipse depicted by the CCC-based model is more compact and it encloses the samples more tightly; while the SCC-based model looks a little conservative but it approximates the CCC-based model well. Secondly, for the cases when experimental samples do not distribute quite uniformly within a bounded connected domain, the SCC-based model may not be able to include those marginal samples in. As shown in Fig. 10(a), all of the samples are required to be enclosed for evaluation of the CCC; this is determined by the definition of CCC. Different with CCC, SCC cares more about the correlation information of the majority of the samples, thus the ellipse it creates may not ensure to enclose those extremely marginal experimental samples, as shown in Fig. 10(b).

# 3 Model evaluation criteria and assessment indexes

The validity verification of a newly proposed modelling method is generally required before it is applied to practical engineering problems. After that, when the convex models are applied to practical applications, the assessment of the above ME and MP models for a specific uncertain problem is also necessary. Therefore in this section, a theoretic evaluation criterion for validity verification and two model assessment indexes for practical applications are proposed. Specifically, from the theoretical analysis level, the concept of "unbiasedness" is proposed, which can be regarded as a test standard for newly developed modelling methods. From the practical application level, the index "fitness" is proposed to quantify adaptability of a constructed convex model for a specific uncertain problem with given experimental samples; besides, the "volume ratio" reflecting the volume of the constructed uncertainty domain is also provided.

### 3.1 A theoretic evaluation criterion

For convex model approach, the uncertainty domain of the parameters is assumed to be a convex set, and the construction of the convex model is carried out based on this assumption. Then a question arises: if the uncertainty domain of $\mathbf{X} = [X_1, X_2, \cdots, X_n]^T$ is really a convex set denoted as $\Omega_X$, could the convex modelling approach exactly construct the same domain with only the marginal intervals of the uncertain parameters and the samples generated from this set? To answer this question, we propose a concept of "unbiasedness" to verify the above different construction methods in convex modelling. More importantly, this concept can also be used as an evaluation criterion for newly proposed convex modelling methods in future.

**Definition 6.** Denote $\Omega_X$ as the actual uncertainty domain of the uncertain parameters $\mathbf{X} = [X_1, X_2, \cdots, X_n]^T$. Given the predefined marginal intervals of the parameters and a set of samples generated from the domain $\Omega_X$, a set $\hat{\Omega}_X$ can be constructed using a specific convex modelling method. The convex modelling method is called as *unbiased*, if the constructed domain $\hat{\Omega}_X$ by using this method is accordant with the actual domain $\Omega_X$ when the number of the extracted samples is large enough. And we could say such a convex modelling method possesses the *unbiasedness* property.

From Definition 6 we know that, "unbiasedness" is a property which evaluates the ability of a convex modelling method to restore the actual uncertainty domain. Through analysis, we summarize the unbiasedness properties of the convex modelling methods discussed in this paper as shown in Table 1. It can be observed that, with regard to ME model, the uncertainty modelling methods based on CCC or SCC are both unbiased; while for MP-I model, they are both biased. From the penultimate column we see that the CCC-based construction method is biased for all of the MP models, while from the last column it can be found that the SCC-based construction method is unbiased for all the convex models except for the MP-I model. Proofs are given in Appendix B. It should be pointed out that although in the proofs the uncertain parameters are assumed to be uniformly distributed in the convex set when evaluating the unbiasedness of the SCC-based method, as mentioned before, it is just a sufficient condition rather than a necessary one. If the samples are sufficient and uniformly scattered within the uncertainty domain, from Appendix B and Table 1 we know that an unbiased convex modelling method can restore this uncertainty domain completely. In practical problems, the experimental samples may not be sufficient or may not be uniformly scattered, and in these cases the presented unbiased convex modelling methods generally can provide an acceptable approximate solution due to the unbiasedness property.

To certain extent, the proposed unbiasedness property actually provides a mathematical criterion for validity verification of a convex modelling method, from the theoretic analysis level. By Table 1, the following three presumptions can be made. Firstly, to construct a convex model through the SCC-based correlation analysis technique will theoretically perform better than the CCC-based ones if the samples are relatively uniform. Secondly, the ME model may show better properties in application because that both the CCC and the SCC-based methods can lead to unbiased results. Thirdly, the MP-I model may not work well for practical problems since it does not satisfy the unbiasedness property no matter which kind of correlation coefficient is used.

### 3.2 Model assessment indexes

#### 3.2.1 Fitness

Given a group of uncertain parameters with experimental samples, different convex models can be constructed to depict the uncertainty domain. For different kinds of uncertain problems, the convex models may present with different adaptabilities, and it is possible that not all these models can perform well for the uncertain parameters with given samples. Therefore, to quantify the fitting degree of a convex model to the given samples, the concept of "fitness" is suggested for model assessment in convex modelling, by which we aim to find those convex uncertainty domains enclosing as many samples as possible.

**Definition 7.** Denote $\Omega_X$ as a constructed uncertainty domain of the parameters $\mathbf{X} = [X_1, X_2, \cdots, X_n]^\mathrm{T}$ with experimental samples $\mathbf{x}^{(s)}$, $s = 1, 2, \cdots, N_s$. If $N_s^*$ samples among the total $N_s$ samples are not enclosed within this domain $\Omega_X$, the *fitness* can be then defined as

$$\kappa = \frac{N_s - N_s^*}{N_s} \tag{28}$$

For practical problems, with a certain group of samples provided, the fitness could tell us that whether the constructed convex model can enclose all the given samples or not. When the constructed convex set can enclose only a minor part of all the samples, it indicates that this constructed convex model may be unsuitable for this group of samples; in this case, another convex model or construction method may need to be tried.

#### 3.2.2 Volume ratio

Traditionally, the best convex model is identified as the one which contains all given experimental samples but has a minimum volume [15]. By our proposed convex modelling methods, the index "fitness" is utilized to judge whether all the samples are enclosed within the constructed uncertainty domain. Besides this, to have an understanding of the volume of the constructed domains, herein the "volume ratio" is furtherly defined as a reference index for the constructed convex models with satisfactory fitness.

**Definition 8.** Denote $V^*$ as the volume of the constructed uncertainty domain $\Omega_X$, and $V$ as the volume of the hyper-rectangle $\bar{\Omega} = [X_1^L, X_1^U] \times [X_2^L, X_2^U] \times \cdots \times [X_n^L, X_n^U]$ determined by the marginal intervals. The volume ratio $v$ is defined as

$$v = \frac{V^*}{V} \times 100\% \tag{29}$$

Additionally, the standard volume ratio $\bar{v}$ is defined as

$$\bar{v} = \sqrt[n]{v} \tag{30}$$

The volume ratio $v$ can be regarded as a ratio of the valid domain when considering correlation between interval variables to the multidimensional box without consideration of correlation. The standard volume ratio $\bar{v}$ is defined to eliminate the influence caused by the dimension $n$.

In this paper, the ME model and the MP models to be constructed can be expressed by Eq. (14) and by Eq. (20), respectively. The volume of the hyper-rectangle $\bar{\Omega}$ is

$$V = 2^n \det(\mathbf{D}_X) \tag{31}$$

For the ME model, the volume of the uncertainty domain expressed by Eq. (14) is

$$V^* = V_E^* = A\sqrt{|\det(\mathbf{C}_X)|} = A\sqrt{|\det(\mathbf{R})|}\det(\mathbf{D}_X) \tag{32}$$

in which $A$ is the volume of an $n$-dimensional standard sphere with a unit radius [18]

$$A = \frac{\pi^{\frac{n}{2}}}{\Gamma\left(\frac{n+2}{2}\right)} \tag{33}$$

where $\Gamma(n) = (n-1)!$ denotes the Gamma function. For the MP models, the volume of the uncertainty domain expressed by Eq. (20) is

$$V^* = V_P^* = 2^n |\det(\mathbf{D}_X \mathbf{S})| = 2^n |\det(\mathbf{S})| \det(\mathbf{D}_X) \tag{34}$$

## 4 Numerical examples and discussions

In this section, four examples on construction of different convex models are investigated. In the first one, the convex models for three standard interval variables with hypothetical samples are constructed and compared. In the second one, a geotechnical engineering problem with uncertain material parameters is considered, and both the ME and the MP-II models are applied for uncertainty quantification of uncertain parameters. Finally in the last two examples, the proposed models are applied in the stress analysis of a cantilever beam and the thermal analysis of augmented reality glasses.

### 4.1 A three-dimensional problem

In this example, three uncertain parameters $X_1$, $X_2$ and $X_3$ with 20 groups of hypothetical samples are assumed to have been regularized to standard interval variables $U_1$, $U_2$ and $U_3$; corresponding transformed samples are listed in Table 2. Since that the uncertainty domain $\Omega_X$ of $\mathbf{X} = [X_1, X_2, X_3]^T$ can be easily obtained from that of $\mathbf{U} = [U_1, U_2, U_3]^T$ by use of the transformation in Eq. (7), in the following the construction of uncertainty domain $\Omega_U$ of **U** is focused on. Different convex models are utilized for quantification of the uncertainty. In the first part, we consider to use the CCC in the creation of the correlation matrix, and subsequently formulating the mathematical expression of $\Omega_U$. In the second part, the constructions are performed based on the SCC-formulated correlation matrix.

**4.1.1 Construction based on CCC-formulated correlation matrix**

Firstly, the ME and the MP models are constructed based on the CCC-formulated correlation matrices. The results and the 3D figures are presented.

**a) Multidimensional Ellipsoid model-based uncertainty quantification**

To create the correlation matrix, three ellipses that quantify the correlation between each pair of interval variables are obtained. By definition of the CCC in ME model as Eq. (9), the CCCs are calculated as shown in Fig. 11. Then the correlation matrix can be obtained as

$$\mathbf{R} = \begin{bmatrix} 1 & 0.7623 & -0.8831 \\ 0.7623 & 1 & -0.6732 \\ -0.8831 & -0.6732 & 1 \end{bmatrix} \quad (35)$$

Furthermore, the analytical expression of the ellipsoid-shaped uncertainty domain $\Omega_U$ of **U** can be obtained by Eq. (12)

$$\Omega_U = \left\{ \mathbf{U} \middle| \begin{bmatrix} U_1 & U_2 & U_3 \end{bmatrix} \begin{bmatrix} 1 & 0.7623 & -0.8831 \\ 0.7623 & 1 & -0.6732 \\ -0.8831 & -0.6732 & 1 \end{bmatrix}^{-1} \begin{bmatrix} U_1 \\ U_2 \\ U_3 \end{bmatrix} \leq \begin{bmatrix} 1 \\ 1 \\ 1 \end{bmatrix} \right\} \quad (36)$$

In Fig. 12, the 3D figure of the constructed ellipsoid-shaped domain and its projections on different planes are presented. The projections are ellipses, and they reflect the correlations between the uncertain parameters $U_1$ and $U_2$, $U_1$ and $U_3$, $U_2$ and $U_3$, respectively. From Lemma 2 we can also get the analytical expressions of these projections as in the following, which is accordant with the ones created for correlation evaluation as shown in Fig. 11. In Fig. 12, the sample points are marked by small circles, and those simultaneously marked with stars are samples locate outside the constructed ellipsoid-shaped uncertainty domain. In this case, totally 18 samples among 20 are enclosed within the constructed uncertainty domain, hence the fitness is $\kappa = 18/20$. The volume ratio is $\nu = 15.90\%$, indicating that the volume of the ellipsoid takes only 15.90% of that of its circumscribed cube $\bar{\Omega} = [-1,1] \times [-1,1] \times [-1,1]$; neglecting the influence of the dimension, the standard volume ratio is derived as $\bar{\nu} = 54.18\%$.

$$\begin{aligned} \begin{bmatrix} U_1 & U_2 \end{bmatrix} \begin{bmatrix} 1 & 0.7623 \\ 0.7623 & 1 \end{bmatrix}^{-1} \begin{bmatrix} U_1 \\ U_2 \end{bmatrix} \leq 1 & \quad \textit{projection on } U_1 - U_2 \textit{ plane} \\ \begin{bmatrix} U_1 & U_3 \end{bmatrix} \begin{bmatrix} 1 & -0.8831 \\ -0.8831 & 1 \end{bmatrix}^{-1} \begin{bmatrix} U_1 \\ U_3 \end{bmatrix} \leq 1 & \quad \textit{projection on } U_1 - U_3 \textit{ plane} \quad (37) \\ \begin{bmatrix} U_2 & U_3 \end{bmatrix} \begin{bmatrix} 1 & -0.6732 \\ -0.6732 & 1 \end{bmatrix}^{-1} \begin{bmatrix} U_2 \\ U_3 \end{bmatrix} \leq 1 & \quad \textit{projection on } U_2 - U_3 \textit{ plane} \end{aligned}$$

**b) Multidimensional Parallelepiped model-based uncertainty quantification**

In this part, different MP models are utilized for convex modelling, through the CCC-formulated correlation matrix.

*(5) MP-II model*

Similar to the construction of the ME model, as shown in Fig. 13, the CCCs between the standard interval variables defined by the MP-II model are firstly calculated by Eq. (9), with which the correlation matrix $\mathbf{R}$, the CSM $\mathbf{H}$ and the Shape Matrix $\mathbf{S}$ can be created as

$$\mathbf{R} = \begin{bmatrix} 1 & 0.73 & -0.86 \\ 0.73 & 1 & -0.58 \\ -0.86 & -0.58 & 1 \end{bmatrix}, \mathbf{H} = \begin{bmatrix} 0.81 & 0.37 & -0.46 \\ 0.37 & 0.90 & -0.24 \\ -0.46 & -0.24 & 0.86 \end{bmatrix}, \mathbf{S} = \begin{bmatrix} 0.49 & 0.22 & -0.28 \\ 0.24 & 0.60 & -0.15 \\ -0.30 & -0.15 & 0.55 \end{bmatrix} \quad (38)$$

The mathematical expression of the parallelepiped-shaped uncertainty domain $\Omega_U$ can be then formulated by Eq. (16) as

$$\Omega_U = \left\{ \mathbf{U} \left| \left| \begin{bmatrix} 0.4939 & 0.2232 & -0.2830 \\ 0.2432 & 0.6000 & -0.1568 \\ -0.2981 & -0.1516 & 0.5504 \end{bmatrix}^{-1} \begin{bmatrix} U_1 \\ U_2 \\ U_3 \end{bmatrix} \right| \leq \begin{bmatrix} 1 \\ 1 \\ 1 \end{bmatrix} \right\} \quad (39)$$

The 3D uncertainty domain is plotted in Fig. 14. And the fitness is obtained as $\kappa = 17/20$, indicating that totally 17 samples among the 20 are enclosed within the constructed domain; the volume ratio is $v = 9.17\%$, and the standard volume ratio is $\bar{v} = 45.10\%$.

In the same way, the other MP models can also be efficiently constructed. The constructed 3D uncertainty domains and their 2D projections are provided in Fig. 15. To some degree, the projections can be recognized as good reflections of the correlations between uncertain parameters. Information on the fitness, the volume ratio and the standard volume ratio is listed in Table 3. From the results in Table 3 it can be seen that none of the convex models constructed through the CCCs can enclose all the given samples.

**4.1.2 Construction based on SCC-formulated correlation matrix**

In the above section, the convex models are constructed through the CCC-formulated correlation matrix. As explained previously, the SCC is also an optional tool for correlation quantification by which the correlation matrix can be created in the same way. Subsequently, the mathematical expression of the convex uncertainty domain of the interval variables can be formulated. Regardless of which type of convex model will be selected, the SCC-formulated correlation matrix is unified since that the SCCs are derived from the samples directly. From the samples listed in Table 2, the correlation matrix can be obtained as

$$\mathbf{R} = \begin{bmatrix} 1 & 0.6361 & -0.7102 \\ 0.6361 & 1 & -0.3422 \\ -0.7102 & -0.3422 & 1 \end{bmatrix} \quad (40)$$

With the correlation matrix $\mathbf{R}$, the uncertainty domain $\Omega_U$ of the standard interval variables $\mathbf{U}$ can be constructed by following Eq. (12) (for ME model) or Eq. (16) (for MP model). The uncertainty domains depicted by different convex models are presented in Fig. 16; and relevant index results are listed in Table 4. From figures in Fig. 16 or the fitness values listed in the second column of Table 4. It can be found that firstly, the ME model, the MP-II model and the UTri-MP model constructed though the SCC can enclose all of the given samples; secondly, for these convex models enclosing all the samples, their volumes differs a lot. Such as for ME model and MP-II model, the volume ratios are 27.86% and 17.33% respectively, indicating that the volume of the ellipsoid-shaped convex uncertainty domain is much larger than the one depicted by the MP-II convex domain. Therefore, it is likely that in this case the MP-II convex model will be more accurate and suitable to quantify the uncertainty of the uncertain parameters $U_1$, $U_2$ and $U_3$. Furthermore, it is also observed that the MP-I model constructed through the SCC performs badly for the given samples, because only 5 among the 20 samples are enclosed within the constructed convex set. This phenomenon is in fact caused by the biasedness mathematical property of the MP-I model constructed through the SCC-formulated correlation matrix, as illustrated in Table 1.

By comparing the performances of the CCC- and the SCC-based models constructed in this example, one can find that the SCC-based ones adapts the given group of samples better than the CCC-based models; however, except for the MP-I model. Indeed, for most practical problems, the SCC-based models exhibit better adaptabilities. Nevertheless, we cannot make the conclusion that the SCC-based model is absolutely better, since that there are always different circumstances among the complicated engineering problems; for example, the geotechnical engineering problem as presented in the following section.

## 4.2 A geotechnical uncertainty problem

The material parameters of rock and soil masses are often variables with uncertainty. They are mutually dependent because of their interaction between each other. Recognition and quantification of the correlation between these physical and mechanical parameters of the rock and soil masses have important theoretical and practical significance in geotechnical engineering. However, as we all know, the experimental samples of geotechnical parameters are rare and costly to obtain in practical engineering. Therefore, in general it is difficult to quantify the correlation and uncertainty of the material parameters with very limited samples by traditional statistical or probabilistic techniques.

Researches indicate that, the residual strength of the argillation intermediate layer such as shale and marlstone in karst region generally has strong dependence on parameters such as clay content, liquid limit, plasticity index, specific surface area, carbonate content, etc. The experimental samples are obtained as listed in Table 5 [44]. By theory of geotechnical engineering we know that some of the parameters can have obvious cross influences on each other. For lack of enough experimental samples, in this example we expect to quantify the correlation between the material uncertain parameters by the proposed convex modelling method, thus providing a quantitative description for their uncertainty domain. The variation intervals of these parameters are set as listed in Table 6.

The ME and the MP-II models are constructed for description of the uncertainty domain for these uncertain parameters; both the CCC and the SCC are considered for creation of the correlation matrix. The results are obtained as shown in Table 7. From Table 7 we known that the ME model with CCC-formulated correlation matrix adapts this problem well, because that all of the 10 experimental samples can be enclosed within the constructed ME-shaped domain. The correlation matrix $\mathbf{R}$ of the uncertain variables $\mathbf{X} = [X_1 \ X_2 \ X_3 \ X_4 \ X_5 \ Y]^\mathrm{T}$ formulated by the CCCs is

$$\mathbf{R} = \begin{bmatrix} 1 & 0.8278 & 0.8314 & 0.5534 & -0.2433 & -0.8100 \\ 0.8278 & 1 & 0.9472 & 0.8171 & 0.4179 & -0.7280 \\ 0.8314 & 0.9472 & 1 & 0.8332 & 0.3573 & -0.9080 \\ 0.5534 & 0.8171 & 0.8332 & 1 & -0.9080 & -0.5712 \\ -0.2433 & 0.4179 & 0.3573 & -0.9080 & 1 & -0.2682 \\ -0.8100 & -0.7280 & -0.9080 & -0.5712 & -0.2682 & 1 \end{bmatrix} \quad (41)$$

and the mathematical expression of the ME-shaped uncertainty domain $\Omega_X$ can be then derived as

$$\Omega_X = \left\{ \mathbf{X} \middle| (\mathbf{X} - \mathbf{X}^m)^\mathrm{T} \begin{bmatrix} 361 & 299 & 190 & 1262 & -231 & -2.31 \\ 299 & 361 & 216 & 1863 & 397 & -2.08 \\ 190 & 216 & 144 & 1200 & 214 & -1.63 \\ 1262 & 1863 & 1200 & 14400 & -5448 & -10.3 \\ -231 & 397 & 214 & -5448 & 2500 & -2.01 \\ -2.31 & -2.08 & -1.63 & -10.3 & -2.01 & 0.0225 \end{bmatrix}^{-1} (\mathbf{X} - \mathbf{X}^m) \leq 1 \right\} \quad (42)$$

in which $\mathbf{X}^m = [47 \ 31 \ 14 \ 120 \ 50 \ 0.45]^\mathrm{T}$ is the midpoint vector. From the correlation matrix we can have a general overview of the dependency degree of these material parameters. For example, in the last column of $\mathbf{R}$, the CCCs indicate the dependency degree between the residual strength and other material parameters; and the values demonstrate that the residual strength is negatively correlated with the other material parameters. By comparison of the absolute values of the CCCs, it can be concluded that the residual strength is relatively strong correlated with the plasticity index $X_3$; while it is relatively weak correlated with the carbonate content $X_5$, negatively. These results agree with that of the qualitative analysis to a certain extent, and illustrating the effectiveness of our proposed convex modelling methods vigorously.

From Table 7 it can also be observed that for this practical problem, the uncertainty domains depicted by the MP-II model and the SCC-based ME model are not able to provide satisfactory solutions. Both the CCC and the SCC-based MP-II models can enclose none of the 10 experimental samples, which are extremely unacceptable in this case. And even though the CCC-based ME model performs very well for this problem, the ME model constructed through the SCC fits the samples badly. Therefore, in practical applications, it cannot be recognized that some of the convex models are definitely better than the others. For specific problems, the most appropriate model should be adopted by using the model assessment indexes.

### 4.3 A cantilever beam problem

In this example the quantification of the uncertain parameters and the non-probabilistic reliability analysis of a cantilever beam as shown in Fig. 17 are investigated [18]. The maximum stress at the fixed end of the cantilever should be less than a yield strength $S$, and the limit-state function can be expressed in the following form

$$g(b,h,L) = S - \frac{6P_x L}{b^2 h} - \frac{6P_y L}{bh^2} \qquad (43)$$

Where $L$ is the length of the beam; $b$ and $h$ are the width and height of the cross section, respectively; $P_x=50,000$N and $P_y=25,000$N are the horizontal force and vertical force applied to the beam, respectively. In this problem, the geometric parameters $b$, $h$ and $L$ are modelled as interval variables with variation intervals $b^I = [90,110]$mm, $h^I = [180,220]$mm and $L^I = [900,1100]$mm. Herein we construct the uncertainty domain of the uncertain parameters with their 32 samples listed in Table 8 by the SCC-based ME and MP-II models proposed in this paper. The SCC-formulated correlation matrix $\mathbf{R}$ is created as

$$\mathbf{R} = \begin{bmatrix} 1 & 0.0342 & 0.3011 \\ 0.0342 & 1 & -0.0019 \\ 0.3011 & -0.0019 & 1 \end{bmatrix} \qquad (44)$$

Consequently, the uncertainty domains depicted by the ME model and the MP-II model can be analytically expressed as follows

a) depicted by the ME model

$$\begin{bmatrix} (b-100)/10 \\ (h-200)/20 \\ (L-1000)/100 \end{bmatrix}^T \begin{bmatrix} 1 & 0.0342 & 0.3011 \\ 0.0342 & 1 & -0.0019 \\ 0.3011 & -0.0019 & 1 \end{bmatrix}^{-1} \begin{bmatrix} (b-100)/10 \\ (h-200)/20 \\ (L-1000)/100 \end{bmatrix} \leq 1 \qquad (45)$$

b) depicted by the MP-II model

$$\left| \begin{bmatrix} 0.8534 & 0.0150 & 0.1316 \\ 0.0171 & 0.9807 & -0.0023 \\ 0.1333 & -0.0020 & 0.8647 \end{bmatrix}^{-1} \begin{bmatrix} (b-100)/10 \\ (h-200)/20 \\ (L-1000)/100 \end{bmatrix} \right| \leq \begin{bmatrix} 1 \\ 1 \\ 1 \end{bmatrix} \qquad (46)$$

The fitness of the constructed ME model and the MP-II model in this case are $\kappa_E = 32/32$ and $\kappa_P = 32/32$ respectively, indicating that both the constructed ME model and the MP-II model can quantify the parametric uncertainty appropriately. The volume ratios of the constructed domains are $v_E = 49.90\%$ and $v_P = 70.62\%$, respectively. This shows that firstly, the valid domain of uncertainty $\Omega$ takes a relatively small part of the box $\bar{\Omega}$ determined by independency assumption of the interval variables, illustrating again the necessity of consideration of the parameter correlation. Secondly, it can also be recognized that the ME model is more adaptable to the geometric parameters in this problem, since that the ellipsoid-shaped domain is smaller in volume than the parallelepiped. With the ME or the MP-II convex uncertainty domain obtained, subsequent uncertainty analyses such as non-probabilistic reliability analysis [16, 18, 28, 29] can be then carried out. Herein we employ the non-probabilistic reliability index to evaluate the reliability degree of a structure under uncertainty, which represents a minimal distance from the original point to the limit-state surface in a standardized space [16, 29]. Note that the distance metric in the corresponding standardized space for the ME model and the MP model are defined by the Euclidean norm and the infinity norm respectively. For this cantilever beam stress analysis, take the constructed ME uncertainty domain as an example, the non-probabilistic reliability index is evaluated by the method given in Ref. [16, 28]. The non-probabilistic reliability index of the stress of this cantilever beam is computed as $\eta_E = 2.0$, indicating safety of the beam structure subjected to the loads in all uncertain cases of the sizes.

### 4.4 An augmented reality glasses problem

Augmented reality glasses (AR glasses) [45, 46] is a kind of smart wearable device developed recently. Because it integrates the computing, communication, positioning, photography and many other functions, this kind of high-tech equipment exhibits essential application potential in different critical areas such as education, healthcare, security and aerospace etc. Similar to the other smart wearable devices such as the smart bracelet and the balance bike, various design requirements need to be considered in the structural design process of the AR glasses, especially for the comfortability and safety.

In this example, the AR glasses as shown in Fig. 18(a) is composed of 5 components, including spectacle frame, micro projector, micro camera, controller and cell. Among them, the light weight design and heat dissipation design of the controller play an important role on the wearing comfort and the operational safety of the device. The exploded diagram of the controller is presented in Fig. 18(b). Because that the AR glasses often operates under uncertain conditions, such as changing environmental temperatures and different power dissipation, the practical operation temperature of the controller generally exhibits high uncertainty. Herein in this example the environmental temperature $T_a$, the air velocity $V_a$, the power dissipation of chip A and chip B denoted as $P_A$ and $P_B$ are modelled as interval variables. The variation intervals are listed in Table 9, and totally 56 groups of simulation samples are utilized. The uncertainty domain of these parameters constructed by SCC-based MP-II convex model can be expressed as

$$\Omega = \left\{ \begin{bmatrix} T_a \\ V_a \\ P_A \\ P_B \end{bmatrix} \middle| \left\| \begin{bmatrix} 24.4119 & 1.0933 & -2.7287 & -1.7661 \\ 0.0127 & 0.2828 & -0.0234 & -0.0310 \\ -0.0216 & -0.0160 & 0.1686 & 0.0939 \\ -0.0047 & -0.0071 & 0.0315 & 0.0567 \end{bmatrix}^{-1} \begin{bmatrix} T_a - 20 \\ V_a - 0.35 \\ P_A - 0.3 \\ P_B - 0.1 \end{bmatrix} \right\| \leq \begin{bmatrix} 1 \\ 1 \\ 1 \\ 1 \end{bmatrix} \right\} \quad (47)$$

The fitness of the constructed model is $\kappa = 54/56$, indicating an acceptable convex modelling for uncertainty of the parameters.

To guarantee the operating safety of the user and the normal operation of the AR glasses, the temperatures of the controller at region A and region B are required to be under a certain value. Therefore herein the non-probabilistic reliability analysis on the temperature of the controller is carried out for the AR glasses. The numerical simulation model of the controller is presented in Fig. 19; figure 19(a) shows the surface temperature of the shell by a sampling simulation, and Fig. 19(b) shows the temperature of the circuit board. The simulation model contains 4 parts, 22928 thermal coupling eight-node hexahedrons. In order to improve efficiency for subsequent non-probabilistic reliability analysis, an accurate second order polynomial response surface was established for the temperatures of the controller at region A and region B; the temperatures are denoted as $T_A$ and $T_B$ respectively

$$\begin{aligned}T_A =\ &100.47 - 124.32d_1 - 14.39d_2 + 10.74d_3 + 1.50d_1d_2 + 10.10d_1d_3 + 0.29d_2d_3 + 38.32d_1^2 + 4.28d_2^2 - 8.9d_3^2 \\ &+ 13.65P_A - 42.59P_B + 17.51P_AP_B + 4.85P_A^2 + 145.24P_B^2 + 0.90T_a - 34.41V_a - 0.18T_aV_a + 28.41V_a^2 \\ T_B =\ &73.27 - 88.5d_1 - 11.92d_2 + 10.13d_3 + 0.54d_1d_2 + 6.67d_1d_3 + 1.15d_2d_3 + 27.81d_1^2 + 3.45d_2^2 - 7.51d_3^2 \\ &+ 0.92T_a - 34.73V_a - 0.09T_aV_a + 32.49V_a^2 + 6.60P_A - 28.69P_B + 26.31P_AP_B + 3.63P_A^2 + 81.92P_B^2\end{aligned} \quad (48)$$

In the above temperature response functions [46], $d_i, i = 1, 2, 3$ represent the structural sizes of the shell of the controller, which are treated as deterministic values in this example. Considering that the allowable values of $T_A$ and $T_B$ are $39\,°C$ and $42\,°C$ respectively, the non-probabilistic indexes for evaluation of the reliability of the operating temperatures are computed, which yields $\eta_A = 1.107$ and $\eta_B = 1.151$. Results show that the operating temperatures of the controller will always meet the temperature requirements since that the non-probabilistic reliability indexes are both greater than 1. If any of the reliability doesn't meet the requirement, the structural parameters such as the sizes and the power dissipation may need to be redesigned.

# 5 Conclusions

In this paper, a unified framework for construction of the non-probabilistic convex models is proposed for uncertainty quantification of structural uncertain parameters. Two kinds of convex models that could take correlation into consideration, namely the multidimensional ellipsoid (ME) model and the multidimensional parallelepiped (MP) model are discussed. In the system of MP model, five different multidimensional parallelepiped-shaped models are included for practical engineering problems with various kinds of uncertainty.

In the construction procedure of a convex model, correlation analysis of interval variables plays a primary role and determines the characteristic matrix in the analytical expression of the uncertainty domain. The correlation coefficient of interval variables can be quantified in two ways, namely, the Convex Correlation Coefficient (CCC) and the Sample Correlation Coefficient (SCC). The CCC has different definitions in different models, and it is evaluated through geometric characteristics of the 2-dimensional convex set that encloses all the bivariate samples. The SCC is defined directly through the bivariate samples and has a united form. In practical, engineers can adopt any type of CCC or SCC for formulation of the correlation matrix, and subsequently construct a ME or MP model for uncertainty quantification.

To provide a theoretic evaluation criterion for validity verification of the modelling methods, the concept of "unbiasedness", which can be regarded as a test standard for subsequent newly proposed convex modelling methods is advocated. With the evaluation criterion of "unbiasedness", the validities of the ME model and the MP models constructed through the CCC and the SCC-formulated correlation matrices are verified. Through theoretic analysis, it is found that the ME model possesses better property, while the MP-I model is proved to be an invalid convex model.

To evaluate the applicability of convex models to a specific problem, two model assessment indexes, namely, the "fitness" and the "volume ratio", are also given. The index "fitness" is used to quantify the adaptability of a constructed convex model for a specific uncertain problem with given experimental samples, by which engineers can judge how well the model fits for the samples. The "volume ratio" is provided as a reference index that reflects the volume of the constructed uncertainty domain.

In numerical examples, the convex models are constructed and compared for uncertainty quantification. As illustrated by our theoretic analysis, it is verified that for general cases the SCC-based convex models can perform better than the CCC-based ones. But it is not absolute in all circumstances; as an exceptional case, the geotechnical engineering problem with uncertain material parameters is also given, for which only the CCC-based ME model can fit the parametric uncertainty well. Therefore, in practical we suggest to select the most appropriate model for a specific problem, according to the indexes such as "fitness" and "volume ratio". Finally, the convex modelling approach and subsequent non-probabilistic reliability analysis are successfully applied to the stress analysis of a cantilever beam under geometrical uncertainties and the temperature response analysis of a pair of AR glasses under operation condition uncertainties.

# Acknowledgements

This work is supported by the National Science Fund for Distinguished Young Scholars (Grant No. 51725502), the Major Program of National Natural Science Foundation of China (Grant No. 51490662), the National Key Research and Development Project of China (Grant No. 2016YFD0701105), the Fok Ying-Tong Education Foundation, China (Grant No. 131005), the Open Funds for State Key Laboratory of Advanced Design and Manufacturing for Vehicle Body, China (31515010).

# Appendix

**Appendix A**

**Proof.** The uncertainty domain expressed by Eq. (12) is equivalent to the following form

$$\Omega_U = \left\{ \tilde{\mathbf{U}} \middle| \tilde{\mathbf{U}}^{\mathrm{T}} \begin{bmatrix} r_{ii} & r_{ij} & r_{i1} & \cdots & r_{in} \\ r_{ji} & r_{jj} & r_{j1} & \cdots & r_{jn} \\ r_{1i} & r_{1j} & r_{11} & \cdots & r_{1n} \\ \vdots & \vdots & \vdots & \vdots \\ r_{ni} & r_{nj} & r_{n1} & \cdots & r_{nn} \end{bmatrix}^{-1} \tilde{\mathbf{U}} \leq 1 \right\} \quad (A.1)$$

in which $\tilde{\mathbf{U}} = [U_i \ U_j \ U_1 \ \cdots \ U_{i-1} \ U_{i+1} \ \cdots \ U_{j-1} \ U_{j+1} \ \cdots \ U_n]^{\mathrm{T}}$ is just a rearranged form of $\mathbf{U}$ by bringing $U_i$ and $U_j$ to the front. Correspondingly, denote the above rearranged matrix as $\tilde{\mathbf{R}}$ and partition it into 4 blocks as

$$\tilde{\mathbf{R}} = \begin{bmatrix} r_{ii} & r_{ij} & r_{i1} & \cdots & r_{in} \\ r_{ji} & r_{jj} & r_{j1} & \cdots & r_{jn} \\ r_{1i} & r_{1j} & r_{11} & \cdots & r_{1n} \\ \vdots & \vdots & \vdots & \vdots & \vdots \\ r_{ni} & r_{nj} & r_{n1} & \cdots & r_{nn} \end{bmatrix} = \begin{bmatrix} \tilde{\mathbf{R}}_B & \tilde{\mathbf{R}}_A^\mathrm{T} \\ \tilde{\mathbf{R}}_A & \tilde{\mathbf{R}}_D \end{bmatrix} \qquad (A.2)$$

Denote the inverse matrix of $\tilde{\mathbf{R}}$ as $\tilde{\mathbf{G}}$ and also partition it into 4 blocks as

$$\tilde{\mathbf{G}} = \tilde{\mathbf{R}}^{-1} = \begin{bmatrix} \tilde{\mathbf{G}}_B & \tilde{\mathbf{G}}_A^\mathrm{T} \\ \tilde{\mathbf{G}}_A & \tilde{\mathbf{G}}_D \end{bmatrix} \qquad (A.3)$$

The matrix $\tilde{\mathbf{G}}$ is symmetrical because $\tilde{\mathbf{R}}$ is symmetrical. The submatrices $\tilde{\mathbf{G}}_B$, $\tilde{\mathbf{G}}_A$ and $\tilde{\mathbf{G}}_D$ are of same dimensions with $\tilde{\mathbf{R}}_B$, $\tilde{\mathbf{R}}_A$ and $\tilde{\mathbf{R}}_D$, respectively. Then the expression of the $n$-dimensional ellipsoid-shaped uncertainty domain $\Omega_U$ can be rewritten as

$$\Omega_U = \left\{ \tilde{\mathbf{U}} \,\middle|\, \tilde{\mathbf{U}}^\mathrm{T} \begin{bmatrix} \tilde{\mathbf{G}}_B & \tilde{\mathbf{G}}_A^\mathrm{T} \\ \tilde{\mathbf{G}}_A & \tilde{\mathbf{G}}_D \end{bmatrix} \tilde{\mathbf{U}} \leq 1 \right\} \qquad (A.4)$$

Let

$$f = \tilde{\mathbf{U}}^\mathrm{T} \begin{bmatrix} \tilde{\mathbf{G}}_B & \tilde{\mathbf{G}}_A^\mathrm{T} \\ \tilde{\mathbf{G}}_A & \tilde{\mathbf{G}}_D \end{bmatrix} \tilde{\mathbf{U}} - 1 \qquad (A.5)$$

which stands for the surface of the multidimensional ellipsoid domain $\Omega_U$. Considering that the projection of the $n$-dimensional ellipsoid $\Omega_U$ on $U_i - U_j$ plane is an ellipse and can be derived by setting the gradient to zero, namely

$$\left[ \frac{\partial f}{\partial U_1} \cdots \frac{\partial f}{\partial U_{i-1}} \frac{\partial f}{\partial U_{i+1}} \cdots \frac{\partial f}{\partial U_{j-1}} \frac{\partial f}{\partial U_{j+1}} \cdots \frac{\partial f}{\partial U_n} \right]^\mathrm{T} = \mathbf{0}_{(n-2) \times 1} \qquad (A.6)$$

it can be obtained that

$$\begin{bmatrix} \tilde{\mathbf{G}}_A & \tilde{\mathbf{G}}_D \end{bmatrix} \tilde{\mathbf{U}} = \mathbf{0}_{(n-2) \times 1} \qquad (A.7)$$

Equation (A.7) can also be written as

$$\tilde{\mathbf{G}}_A \begin{bmatrix} U_i & U_j \end{bmatrix}^\mathrm{T} + \tilde{\mathbf{G}}_D \tilde{\mathbf{U}}_{\backslash i,j} = \mathbf{0}_{(n-2) \times 1} \qquad (A.8)$$

where $\tilde{\mathbf{U}}_{\backslash i,j} = [U_1 \cdots U_{i-1} \; U_{i+1} \cdots U_{j-1} \; U_{j+1} \cdots U_n]^\mathrm{T}$. From Eq. (A.8) we have

$$\tilde{\mathbf{U}}_{\backslash i,j} = -\tilde{\mathbf{G}}_D^{-1} \tilde{\mathbf{G}}_A \begin{bmatrix} U_i & U_j \end{bmatrix}^\mathrm{T} \qquad (A.9)$$

With Eq. (A.9) there is

$$\tilde{\mathbf{U}} = \begin{bmatrix} \mathbf{E}_2 \\ -\tilde{\mathbf{G}}_D^{-1} \tilde{\mathbf{G}}_A \end{bmatrix} \begin{bmatrix} U_i & U_j \end{bmatrix}^\mathrm{T} \qquad (A.10)$$

in which $\mathbf{E}_2$ is a $2 \times 2$ identity matrix. Substitute Eq. (A.10) into Eq. (A.4), the ellipse-shaped projection of $\Omega_U$ on $U_i - U_j$ plane can be derived as

$$\begin{bmatrix} U_i & U_j \end{bmatrix} \begin{bmatrix} \mathbf{E}_2 \\ -\tilde{\mathbf{G}}_D^{-1}\tilde{\mathbf{G}}_A \end{bmatrix}^T \begin{bmatrix} \tilde{\mathbf{G}}_B & \tilde{\mathbf{G}}_A^T \\ \tilde{\mathbf{G}}_A & \tilde{\mathbf{G}}_D \end{bmatrix} \begin{bmatrix} \mathbf{E}_2 \\ -\tilde{\mathbf{G}}_D^{-1}\tilde{\mathbf{G}}_A \end{bmatrix} \begin{bmatrix} U_i & U_j \end{bmatrix}^T \leq 1 \tag{A.11}$$

in which

$$\begin{bmatrix} \mathbf{E}_2 \\ -\tilde{\mathbf{G}}_D^{-1}\tilde{\mathbf{G}}_A \end{bmatrix}^T \begin{bmatrix} \tilde{\mathbf{G}}_B & \tilde{\mathbf{G}}_A^T \\ \tilde{\mathbf{G}}_A & \tilde{\mathbf{G}}_D \end{bmatrix} \begin{bmatrix} \mathbf{E}_2 \\ -\tilde{\mathbf{G}}_D^{-1}\tilde{\mathbf{G}}_A \end{bmatrix} = \tilde{\mathbf{G}}_B - \tilde{\mathbf{G}}_A^T \tilde{\mathbf{G}}_D^{-1} \tilde{\mathbf{G}}_A \tag{A.12}$$

Denote it as $\mathbf{G}^*$, i.e. $\mathbf{G}^* = \tilde{\mathbf{G}}_B - \tilde{\mathbf{G}}_A^T \tilde{\mathbf{G}}_D^{-1} \tilde{\mathbf{G}}_A$, then the ellipse-shaped projection on $U_i - U_j$ plane expressed by Eq. (A.11) can be rewritten as

$$\begin{bmatrix} U_i & U_j \end{bmatrix} \mathbf{G}^* \begin{bmatrix} U_i & U_j \end{bmatrix}^T \leq 1 \tag{A.13}$$

Because that the matrix $\tilde{\mathbf{G}}$ is the inverse matrix of $\tilde{\mathbf{R}}$, there should be

$$\tilde{\mathbf{G}}\tilde{\mathbf{R}} = \begin{bmatrix} \tilde{\mathbf{G}}_B & \tilde{\mathbf{G}}_A^T \\ \tilde{\mathbf{G}}_A & \tilde{\mathbf{G}}_D \end{bmatrix} \begin{bmatrix} \tilde{\mathbf{R}}_B & \tilde{\mathbf{R}}_A^T \\ \tilde{\mathbf{R}}_A & \tilde{\mathbf{R}}_D \end{bmatrix} = \begin{bmatrix} \tilde{\mathbf{G}}_B\tilde{\mathbf{R}}_B + \tilde{\mathbf{G}}_A^T\tilde{\mathbf{R}}_A & \tilde{\mathbf{G}}_B\tilde{\mathbf{R}}_A^T + \tilde{\mathbf{G}}_A^T\tilde{\mathbf{R}}_D \\ \tilde{\mathbf{G}}_A\tilde{\mathbf{R}}_B + \tilde{\mathbf{G}}_D\tilde{\mathbf{R}}_A & \tilde{\mathbf{G}}_A\tilde{\mathbf{R}}_A^T + \tilde{\mathbf{G}}_D\tilde{\mathbf{R}}_D \end{bmatrix} = \begin{bmatrix} \mathbf{E}_2 & \mathbf{0}_{2\times(n-2)} \\ \mathbf{0}_{(n-2)\times 2} & \mathbf{E}_{n-2} \end{bmatrix} \tag{A.14}$$

from which the equalities as following can be obtained

$$\begin{aligned} \tilde{\mathbf{G}}_B\tilde{\mathbf{R}}_B + \tilde{\mathbf{G}}_A^T\tilde{\mathbf{R}}_A &= \mathbf{E}_2 \\ \tilde{\mathbf{G}}_A\tilde{\mathbf{R}}_B + \tilde{\mathbf{G}}_D\tilde{\mathbf{R}}_A &= \mathbf{0}_{(n-2)\times 2} \end{aligned} \tag{A.15}$$

With Eq. (A.15) we can have

$$\begin{aligned} \mathbf{G}^*\tilde{\mathbf{R}}_B &= \left(\tilde{\mathbf{G}}_B - \tilde{\mathbf{G}}_A^T\tilde{\mathbf{G}}_D^{-1}\tilde{\mathbf{G}}_A\right)\tilde{\mathbf{R}}_B = \tilde{\mathbf{G}}_B\tilde{\mathbf{R}}_B - \tilde{\mathbf{G}}_A^T\tilde{\mathbf{G}}_D^{-1}\tilde{\mathbf{G}}_A\tilde{\mathbf{R}}_B \\ &= \mathbf{E}_2 - \tilde{\mathbf{G}}_A^T\tilde{\mathbf{R}}_A - \tilde{\mathbf{G}}_A^T\tilde{\mathbf{G}}_D^{-1}\left(-\tilde{\mathbf{G}}_D\tilde{\mathbf{R}}_A\right) = \mathbf{E}_2 - \tilde{\mathbf{G}}_A^T\tilde{\mathbf{R}}_A + \tilde{\mathbf{G}}_A^T\tilde{\mathbf{R}}_A \\ &= \mathbf{E}_2 \end{aligned} \tag{A.16}$$

namely

$$\mathbf{G}^* = \tilde{\mathbf{R}}_B^{-1} = \begin{bmatrix} r_{ii} & r_{ij} \\ r_{ji} & r_{jj} \end{bmatrix}^{-1} = \begin{bmatrix} 1 & r_{ij} \\ r_{ji} & 1 \end{bmatrix}^{-1} \tag{A.17}$$

By substituting Eq. (A.17) into Eq. (A.13), the conclusion of Lemma 2 is obtained. □

# Appendix B

## Appendix B1

**Proof.** Assume that the samples of the uncertain parameters $\mathbf{X}^T = \{X_1, X_2, \cdots, X_n\}$ are uniformly distributed within the uncertainty domain $\Omega_X$ that depicted by Eq. (14). Then the Sample Correlation Coefficient (SCC) defined in Eq. (27) can be expressed as

$$r_{s-ij} = \frac{\dfrac{1}{V}\overbrace{\int\cdots\int}^{n}_{\Omega_X}(X_i - X_i^m)(X_j - X_j^m)\mathrm{d}X_1\cdots\mathrm{d}X_n}{\sqrt{\dfrac{1}{V}\overbrace{\int\cdots\int}^{n}_{\Omega_X}(X_i - X_i^m)^2\mathrm{d}X_1\cdots\mathrm{d}X_n}\sqrt{\dfrac{1}{V}\overbrace{\int\cdots\int}^{n}_{\Omega_X}(X_j - X_j^m)^2\mathrm{d}X_1\cdots\mathrm{d}X_n}} \tag{B1.1}$$

or

$$r_{s-ij} = \frac{\frac{1}{V}\overset{n}{\int\cdots\int}_{\Omega_X}\left(\frac{X_i - X_i^m}{X_i^r}\right)\left(\frac{X_j - X_j^m}{X_j^r}\right)dX_1\cdots dX_n}{\sqrt{\frac{1}{V}\overset{n}{\int\cdots\int}_{\Omega_X}\left(\frac{X_i - X_i^m}{X_i^r}\right)^2 dX_1\cdots dX_n}\sqrt{\frac{1}{V}\overset{n}{\int\cdots\int}_{\Omega_X}\left(\frac{X_j - X_j^m}{X_j^r}\right)^2 dX_1\cdots dX_n}} \quad (B1.2)$$

where $V$ is the volume of the $n$-dimensional ellipsoid-shaped uncertainty domain $\Omega_X$, which is also the domain of integration. Let

$$\boldsymbol{\delta} = \mathbf{P}^{-1}\mathbf{D}_X^{-1}(\mathbf{X} - \mathbf{X}^m) \quad (B1.3)$$

in which $\mathbf{P}$ could be any $n$-by-$n$ matrix that satisfies

$$\mathbf{PP}^T = \mathbf{R} \quad (B1.4)$$

Then the domain $\Omega_X$ of integration can be rewritten as

$$\Omega_X = \Omega^* = \{\boldsymbol{\delta} | \boldsymbol{\delta}^T \boldsymbol{\delta} \leq 1\} \quad (B1.5)$$

From Eq. (B1.3) we have

$$\mathbf{D}_X^{-1}(\mathbf{X} - \mathbf{X}^m) = \mathbf{P}\boldsymbol{\delta} \quad (B1.6)$$

and furthermore there is

$$\begin{aligned}\frac{X_i - X_i^m}{X_i^r} &= \mathbf{p}_i \boldsymbol{\delta} \\ \frac{X_j - X_j^m}{X_j^r} &= \mathbf{p}_j \boldsymbol{\delta}\end{aligned} \quad (B1.7)$$

and

$$dX_1 \cdots dX_n = |\det(\mathbf{P})|\det(\mathbf{D}_X)d\delta_1 \cdots d\delta_n \quad (B1.8)$$

where $\mathbf{p}_i$ is the $i$-th row of the matrix $\mathbf{P}$, and $\mathbf{p}_j$ is the $j$-th row vector. Substitute Eq. (B1.7-B1.8) into Eq. (B1.2) there is

$$\begin{aligned}r_{s-ij} &= \frac{\overset{n}{\int\cdots\int}_{\Omega^*}(\mathbf{p}_i\boldsymbol{\delta})(\mathbf{p}_j\boldsymbol{\delta})|\det(\mathbf{P})|\det(\mathbf{D}_X)d\delta_1\cdots d\delta_n}{\sqrt{\overset{n}{\int\cdots\int}_{\Omega^*}(\mathbf{p}_i\boldsymbol{\delta})^2|\det(\mathbf{P})|\det(\mathbf{D}_X)d\delta_1\cdots d\delta_n}\sqrt{\overset{n}{\int\cdots\int}_{\Omega^*}(\mathbf{p}_j\boldsymbol{\delta})^2|\det(\mathbf{P})|\det(\mathbf{D}_X)d\delta_1\cdots d\delta_n}} \\ &= \frac{\overset{n}{\int\cdots\int}_{\Omega^*}(\mathbf{p}_i\boldsymbol{\delta})(\mathbf{p}_j\boldsymbol{\delta})d\delta_1\cdots d\delta_n}{\sqrt{\overset{n}{\int\cdots\int}_{\Omega^*}(\mathbf{p}_i\boldsymbol{\delta})^2 d\delta_1\cdots d\delta_n}\sqrt{\overset{n}{\int\cdots\int}_{\Omega^*}(\mathbf{p}_j\boldsymbol{\delta})^2 d\delta_1\cdots d\delta_n}}\end{aligned}$$

(B1.9)

Because there is

$$(\mathbf{p}_i\boldsymbol{\delta})(\mathbf{p}_j\boldsymbol{\delta}) = (p_{i1}\delta_1 + \cdots + p_{in}\delta_n)(p_{j1}\delta_1 + \cdots + p_{jn}\delta_n) = \left(\sum_{k=1}^{n} p_{ik} p_{jk} \delta_k^2\right) + \left(\sum_{h \neq q}(p_{ih} p_{jq} + p_{jh} p_{iq})\delta_h \delta_q\right)$$

$$(\mathbf{p}_i\boldsymbol{\delta})^2 = (p_{i1}\delta_1 + \cdots + p_{in}\delta_n)^2 = \left(\sum_{k=1}^{n} p_{ik}^2 \delta_k^2\right) + \left(2\sum_{h \neq q} p_{ih} p_{iq} \delta_h \delta_q\right)$$

$$(\mathbf{p}_j\boldsymbol{\delta})^2 = (p_{j1}\delta_1 + \cdots + p_{jn}\delta_n)^2 = \left(\sum_{k=1}^{n} p_{jk}^2 \delta_k^2\right) + \left(2\sum_{h \neq q} p_{jh} p_{jq} \delta_h \delta_q\right)$$

(B1.10)

then it can be obtained that

$$\int_{\Omega^*}\cdots\int (\mathbf{p}_i\boldsymbol{\delta})(\mathbf{p}_j\boldsymbol{\delta}) d\delta_1 \cdots d\delta_n = \int_{\Omega^*}\cdots\int \left(\left(\sum_{k=1}^{n} p_{ik} p_{jk} \delta_k^2\right) + \left(\sum_{h \neq q}(p_{ih} p_{jq} + p_{jh} p_{iq})\delta_h \delta_q\right)\right) d\delta_1 \cdots d\delta_n$$

$$\int_{\Omega^*}\cdots\int (\mathbf{p}_i\boldsymbol{\delta})^2 d\delta_1 \cdots d\delta_n = \int_{\Omega^*}\cdots\int \left(\left(\sum_{k=1}^{n} p_{ik}^2 \delta_k^2\right) + \left(2\sum_{h \neq q} p_{ih} p_{iq} \delta_h \delta_q\right)\right) d\delta_1 \cdots d\delta_n$$

$$\int_{\Omega^*}\cdots\int (\mathbf{p}_j\boldsymbol{\delta})^2 d\delta_1 \cdots d\delta_n = \int_{\Omega^*}\cdots\int \left(\left(\sum_{k=1}^{n} p_{jk}^2 \delta_k^2\right) + \left(2\sum_{h \neq q} p_{jh} p_{jq} \delta_h \delta_q\right)\right) d\delta_1 \cdots d\delta_n$$

(B1.11)

However, since that the domain of integration is symmetric about the origin, there is

$$\int_{\Omega^*}\cdots\int \left(\sum_{h \neq q}(p_{ih} p_{jq} + p_{jh} p_{iq})\delta_h \delta_q\right) d\delta_1 \cdots d\delta_n = 0$$

$$\int_{\Omega^*}\cdots\int \left(2\sum_{h \neq q} p_{ih} p_{iq} \delta_h \delta_q\right) d\delta_1 \cdots d\delta_n = 0 \quad (B1.12)$$

$$\int_{\Omega^*}\cdots\int \left(2\sum_{h \neq q} p_{jh} p_{jq} \delta_h \delta_q\right) d\delta_1 \cdots d\delta_n = 0$$

With Eqs. (B1.10-B1.12), Eq. (1.9) turns to

$$r_{s-ij} = \frac{\int_{\Omega^*}\cdots\int \left(\sum_{k=1}^{n} p_{ik} p_{jk} \delta_k^2\right) d\delta_1 \cdots d\delta_n}{\sqrt{\int_{\Omega^*}\cdots\int \left(\sum_{k=1}^{n} p_{ik}^2 \delta_k^2\right) d\delta_1 \cdots d\delta_n} \sqrt{\int_{\Omega^*}\cdots\int \left(\sum_{k=1}^{n} p_{jk}^2 \delta_k^2\right) d\delta_1 \cdots d\delta_n}}$$

(B1.13)

Additionally

$$\int_{\Omega^*}\cdots\int \left(\sum_{k=1}^{n} p_{ik} p_{jk} \delta_k^2\right) d\delta_1 \cdots d\delta_n = \sum_{k=1}^{n} \int_{\Omega^*}\cdots\int (p_{ik} p_{jk} \delta_k^2) d\delta_1 \cdots d\delta_n$$

$$= \sum_{k=1}^{n} p_{ik} p_{jk} \left(\int_{\Omega^*}\cdots\int (\delta_k^2) d\delta_1 \cdots d\delta_n\right)$$

(B1.14)

Since that the domain of integration is the *n*-dimensional standard sphere with a unit radius, then for

$\forall k$, there is

$$\int\cdots\int_{\Omega^*}^{n} \left(\delta_k^2\right) \mathrm{d}\delta_1 \cdots \mathrm{d}\delta_n = c \tag{B1.15}$$

here $c$ means a constant value. So we have

$$\int\cdots\int_{\Omega^*}^{n} \left(\sum_{k=1}^{n} p_{ik} p_{jk} \delta_k^2\right) \mathrm{d}\delta_1 \cdots \mathrm{d}\delta_n = c\sum_{k=1}^{n} p_{ik} p_{jk} = c\mathbf{p}_i \mathbf{p}_j^{\mathrm{T}} \tag{B1.16}$$

similarly

$$\int\cdots\int_{\Omega^*}^{n} \left(\sum_{k=1}^{n} p_{ik}^2 \delta_k^2\right) \mathrm{d}\delta_1 \cdots \mathrm{d}\delta_n = c\mathbf{p}_i \mathbf{p}_i^{\mathrm{T}}$$

$$\int\cdots\int_{\Omega^*}^{n} \left(\sum_{k=1}^{n} p_{jk}^2 \delta_k^2\right) \mathrm{d}\delta_1 \cdots \mathrm{d}\delta_n = c\mathbf{p}_j \mathbf{p}_j^{\mathrm{T}} \tag{B1.17}$$

Substitute Eqs. (B1.16) and (B1.17) into Eq. (B1.13) there is

$$r_{s-ij} = \frac{\mathbf{p}_i \mathbf{p}_j^{\mathrm{T}}}{\sqrt{\mathbf{p}_i \mathbf{p}_i^{\mathrm{T}}} \sqrt{\mathbf{p}_j \mathbf{p}_j^{\mathrm{T}}}} \tag{B1.18}$$

From Eq. (B1.4) we know furtherly that

$$r_{s-ij} = \frac{r_{ij}}{\sqrt{r_{ii}} \sqrt{r_{jj}}} = r_{ij} \tag{B1.19}$$

Namely, SCC can be accordant with the elements in the assumed matrix $\mathbf{R}$ for the ME model.

For $n=2$, because that the elements in the assumed matrix $\mathbf{R}$ is accordant with CCC, thus there is also SCC=CCC. □

**Appendix B2**

**Proof.** Assume that the samples of the uncertain parameters $\mathbf{X}^{\mathrm{T}} = \{X_1, X_2, \cdots, X_n\}$ are uniformly distributed within the uncertainty domain $\Omega_X$ that depicted by Eq. (20). The formula of the SCC is the same as Eq. (B1. 2), however herein $V$ is the volume of the $n$-dimensional parallelepiped-shaped uncertainty domain $\Omega_X$, which is also the domain of integration. Let

$$\boldsymbol{\delta} = (\mathbf{D}_X \mathbf{S})^{-1}(\mathbf{X} - \mathbf{X}^m) \tag{B2.1}$$

Then the domain $\Omega_X$ of integration can be rewritten as

$$\Omega_X = \Omega^* = \left\{ \boldsymbol{\delta} \middle| [|\boldsymbol{\delta}|] \leq \mathbf{e} \right\} \tag{B2.2}$$

From Eq. (B2.1) we have

$$\mathbf{D}_X^{-1}\left(\mathbf{X} - \mathbf{X}^m\right) = \mathbf{S}\boldsymbol{\delta} \tag{B2.3}$$

and furthermore there is

$$\frac{X_i - X_i^m}{X_i^r} = \mathbf{s}_i \boldsymbol{\delta}$$

$$\frac{X_j - X_j^m}{X_j^r} = \mathbf{s}_j \boldsymbol{\delta}$$

(B2.4)

and

$$\mathrm{d}X_1 \cdots \mathrm{d}X_n = |\det(\mathbf{S})| \det(\mathbf{D}_X) \mathrm{d}\delta_1 \cdots \mathrm{d}\delta_n \qquad (B2.5)$$

where $\mathbf{s}_i$ is the *i*-th row of the Shape Matrix $\mathbf{S}$, and $\mathbf{s}_j$ is the *j*-th row vector. Substitute Eqs. (B2.4-B2.5) into Eq. (B1.2) there is

$$r_{s-ij} = \frac{\int_{\Omega^*}^n \cdots \int (\mathbf{s}_i \boldsymbol{\delta})(\mathbf{s}_j \boldsymbol{\delta}) |\det(\mathbf{S})| \det(\mathbf{D}_X) \mathrm{d}\delta_1 \cdots \mathrm{d}\delta_n}{\sqrt{\int_{\Omega^*}^n \cdots \int (\mathbf{s}_i \boldsymbol{\delta})^2 |\det(\mathbf{S})| \det(\mathbf{D}_X) \mathrm{d}\delta_1 \cdots \mathrm{d}\delta_n} \sqrt{\int_{\Omega^*}^n \cdots \int (\mathbf{s}_j \boldsymbol{\delta})^2 |\det(\mathbf{S})| \det(\mathbf{D}_X) \mathrm{d}\delta_1 \cdots \mathrm{d}\delta_n}}$$

$$= \frac{\int_{\Omega^*}^n \cdots \int (\mathbf{s}_i \boldsymbol{\delta})(\mathbf{s}_j \boldsymbol{\delta}) \mathrm{d}\delta_1 \cdots \mathrm{d}\delta_n}{\sqrt{\int_{\Omega^*}^n \cdots \int (\mathbf{s}_i \boldsymbol{\delta})^2 \mathrm{d}\delta_1 \cdots \mathrm{d}\delta_n} \sqrt{\int_{\Omega^*}^n \cdots \int (\mathbf{s}_j \boldsymbol{\delta})^2 \mathrm{d}\delta_1 \cdots \mathrm{d}\delta_n}}$$

(B2.6)

Because there is

$$(\mathbf{s}_i \boldsymbol{\delta})(\mathbf{s}_j \boldsymbol{\delta}) = (s_{i1}\delta_1 + \cdots + s_{in}\delta_n)(s_{j1}\delta_1 + \cdots + s_{jn}\delta_n)$$
$$= \left(\sum_{k=1}^n s_{ik} s_{jk} \delta_k^2\right) + \left(\sum_{h \neq q} (s_{ih} s_{jq} + s_{jh} s_{iq}) \delta_h \delta_q\right)$$

$$(\mathbf{s}_i \boldsymbol{\delta})^2 = (s_{i1}\delta_1 + \cdots + s_{in}\delta_n)^2 = \left(\sum_{k=1}^n s_{ik}^2 \delta_k^2\right) + \left(2\sum_{h \neq q} s_{ih} s_{iq} \delta_h \delta_q\right)$$

$$(\mathbf{s}_j \boldsymbol{\delta})^2 = (s_{j1}\delta_1 + \cdots + s_{jn}\delta_n)^2 = \left(\sum_{k=1}^n s_{jk}^2 \delta_k^2\right) + \left(2\sum_{h \neq q} s_{jh} s_{jq} \delta_h \delta_q\right)$$

(B2.7)

then it can be obtained that

$$\int_{\Omega^*}^n \cdots \int (\mathbf{s}_i \boldsymbol{\delta})(\mathbf{s}_j \boldsymbol{\delta}) \mathrm{d}\delta_1 \cdots \mathrm{d}\delta_n = \int_{\Omega^*}^n \cdots \int \left[\left(\sum_{k=1}^n s_{ik} s_{jk} \delta_k^2\right) + \left(\sum_{h \neq q} (s_{ih} s_{jq} + s_{jh} s_{iq}) \delta_h \delta_q\right)\right] \mathrm{d}\delta_1 \cdots \mathrm{d}\delta_n$$

$$\int_{\Omega^*}^n \cdots \int (\mathbf{s}_i \boldsymbol{\delta})^2 \mathrm{d}\delta_1 \cdots \mathrm{d}\delta_n = \int_{\Omega^*}^n \cdots \int \left[\left(\sum_{k=1}^n s_{ik}^2 \delta_k^2\right) + \left(2\sum_{h \neq q} s_{ih} s_{iq} \delta_h \delta_q\right)\right] \mathrm{d}\delta_1 \cdots \mathrm{d}\delta_n$$

$$\int_{\Omega^*}^n \cdots \int (\mathbf{s}_j \boldsymbol{\delta})^2 \mathrm{d}\delta_1 \cdots \mathrm{d}\delta_n = \int_{\Omega^*}^n \cdots \int \left[\left(\sum_{k=1}^n s_{jk}^2 \delta_k^2\right) + \left(2\sum_{h \neq q} s_{jh} s_{jq} \delta_h \delta_q\right)\right] \mathrm{d}\delta_1 \cdots \mathrm{d}\delta_n$$

(B2.8)

However, for the domain of integration is symmetric about the origin, then there is

$$\int_{\Omega^*} \cdots \int \left( \sum_{h \neq q} \left( s_{ih} c_{jq} + s_{jh} c_{iq} \right) \delta_h \delta_q \right) d\delta_1 \cdots d\delta_n = 0$$

$$\int_{\Omega^*} \cdots \int \left( 2 \sum_{h \neq q} s_{ih} s_{iq} \delta_h \delta_q \right) d\delta_1 \cdots d\delta_n = 0 \quad \text{(B2.9)}$$

$$\int_{\Omega^*} \cdots \int \left( 2 \sum_{h \neq q} s_{jh} s_{jq} \delta_h \delta_q \right) d\delta_1 \cdots d\delta_n = 0$$

With Eqs. (B2.7-B2.9), Eq. (B2.6) turns to

$$r_{s-ij} = \frac{\int_{\Omega^*} \cdots \int \left( \sum_{k=1}^{n} s_{ik} s_{jk} \delta_k^2 \right) d\delta_1 \cdots d\delta_n}{\sqrt{\int_{\Omega^*} \cdots \int \left( \sum_{k=1}^{n} s_{ik}^2 \delta_k^2 \right) d\delta_1 \cdots d\delta_n} \sqrt{\int_{\Omega^*} \cdots \int \left( \sum_{k=1}^{n} s_{jk}^2 \delta_k^2 \right) d\delta_1 \cdots d\delta_n}} \quad \text{(B2.10)}$$

Additionally

$$\int_{\Omega^*} \cdots \int \left( \sum_{k=1}^{n} s_{ik} s_{jk} \delta_k^2 \right) d\delta_1 \cdots d\delta_n = \sum_{k=1}^{n} \int_{\Omega^*} \cdots \int \left( s_{ik} s_{jk} \delta_k^2 \right) d\delta_1 \cdots d\delta_n$$

$$= \sum_{k=1}^{n} s_{ik} s_{jk} \left( \int_{\Omega^*} \cdots \int \left( \delta_k^2 \right) d\delta_1 \cdots d\delta_n \right) \quad \text{(B2.11)}$$

for $\forall k$, there is

$$\int_{\Omega^*} \cdots \int \left( \delta_k^2 \right) d\delta_1 \cdots d\delta_n = 2^{n-1} \left( \frac{1}{3} \delta_k^3 \Big|_{-1}^{1} \right) = \frac{2^n}{3} \quad \text{(B2.12)}$$

so we have

$$\int_{\Omega^*} \cdots \int \left( \sum_{k=1}^{n} s_{ik} s_{jk} \delta_k^2 \right) d\delta_1 \cdots d\delta_n = \frac{2^n}{3} \sum_{k=1}^{n} s_{ik} c_{jk} = \frac{2^n}{3} \mathbf{s}_i \mathbf{s}_j^{\mathrm{T}} \quad \text{(B2.13)}$$

similarly

$$\int_{\Omega^*} \cdots \int \left( \sum_{k=1}^{n} s_{ik}^2 \delta_k^2 \right) d\delta_1 \cdots d\delta_n = \frac{2^n}{3} \mathbf{s}_i \mathbf{s}_i^{\mathrm{T}}$$

$$\int_{\Omega^*} \cdots \int \left( \sum_{k=1}^{n} s_{jk}^2 \delta_k^2 \right) d\delta_1 \cdots d\delta_n = \frac{2^n}{3} \mathbf{s}_j \mathbf{s}_j^{\mathrm{T}} \quad \text{(B2.14)}$$

Substitute Eqs. (B2.13) and (B2.14) into Eq. (B2.10) there is

$$r_{s-ij} = \frac{\mathbf{s}_i \mathbf{s}_j^{\mathrm{T}}}{\sqrt{\mathbf{s}_i \mathbf{s}_i^{\mathrm{T}}} \sqrt{\mathbf{s}_j \mathbf{s}_j^{\mathrm{T}}}} \quad \text{(B2.15)}$$

From Eq. (17) we know that

$$\mathbf{SS}^{\mathrm{T}} = \mathbf{T}^{-1}\mathbf{H}\left(\mathbf{T}^{-1}\mathbf{H}\right)^{\mathrm{T}} = \mathbf{T}^{-1}\mathbf{H}\mathbf{H}^{\mathrm{T}}\left(\mathbf{T}^{-1}\right)^{\mathrm{T}} \quad (B2.16)$$

For the MP models except the MP-I model, there is

$$\mathbf{R} = \mathbf{H}\mathbf{H}^{\mathrm{T}} \quad (B2.17)$$

Since $\mathbf{s}_i\mathbf{s}_j^{\mathrm{T}}$ is the entry in the $i$-th row and $j$-th column of the matrix $\mathbf{SS}^{\mathrm{T}}$, $\mathbf{s}_i\mathbf{s}_i^{\mathrm{T}}$ ($\mathbf{s}_j\mathbf{s}_j^{\mathrm{T}}$) is the entry in the $i$-th row and $i$-th column (the $j$-th row and $j$-th) of the matrix, the following conclusion can be reached

$$r_{s-ij} = \frac{\mathbf{s}_i\mathbf{s}_j^{\mathrm{T}}}{\sqrt{\mathbf{s}_i\mathbf{s}_i^{\mathrm{T}}}\sqrt{\mathbf{s}_j\mathbf{s}_j^{\mathrm{T}}}} = \frac{(1/w_iw_j)r_{ij}}{\sqrt{(1/w_i^2)r_{ii}}\sqrt{(1/w_j^2)r_{jj}}} = r_{ij} \quad (B2.18)$$

Namely, SCC can be accordant with the elements in the assumed matrix $\mathbf{R}$ for the MP model (except the MP-I model).

For $n=2$, because that the elements in the assumed matrix $\mathbf{R}$ is accordant with CCC, thus there is also SCC=CCC. □

# References

[1] R.G. Sexsmith, Probability-based safety analysis — value and drawbacks, Struct. Saf. 21 (4) (1999) 303-310.
[2] Y. Ben-Haim, I. Elishakoff, Convex Models of Uncertainties in Applied Mechanics, Elsevier Science Publisher, Amsterdam, 1990.
[3] Y. Ben-Haim, Convex models of uncertainty in radial pulse buckling of shells, ASME J. Appl. Mech. 60 (3) (1993) 683.
[4] I. Elishakoff, P. Elisseeff, S. Glegg, Nonprobabilistic, convex-theoretic modelling of scatter in material properties, AIAA J. 32 (4) (1994) 843-849.
[5] Y. Ben-Haim, A non-probabilistic concept of reliability, Struct. Saf. 14 (4) (1994) 227-245.
[6] I. Elishakoff, An idea on the uncertainty triangle, Shock Vib. Digest 22 (10) (1990) 1.
[7] M. Lombardi, R.T. Haftka, Anti-optimization technique for structural design under load uncertainties, Comput. Methods Appl. Mech. Engrg. 157 (1-2) (1998) 19-31.
[8] S. Ganzerli, C.P. Pantelides, Optimum structural design via convex model superposition, Comput. Struct. 74 (6) (2000) 639-647.
[9] L. Fryba, N. Yoshikawa, Bounds analysis of a beam based on the convex model of uncertain foundation, J. Sound Vib. 212 (3) (1998) 547-557.
[10] S. McWilliam, Anti-optimisation of uncertain structures using interval analysis, Comput. Struct. 79 (4) (2001) 421-430.
[11] S.H. Chen, H.D. Lian, X.W. Yang, Interval eigenvalue analysis for structures with interval, Finite Elem. Anal. Des. 39 (5-6) (2003) 419-431.
[12] Z.P. Qiu, I. Elishakoff, Antioptimization of structures with large uncertain-but-non-random parameters via interval analysis. Comput. Methods Appl. Mech. Engrg. 152 (3-4) (1998) 361-372.
[13] A. Sofi, G. Muscolino, I. Elishakoff, Natural frequencies of structures with interval parameters, J. Sound Vib. 347 (2015) 79-95.

[14] D. Moens, D. Vandepitte, Interval sensitivity theory and its application to frequency response envelope analysis of uncertain structures, Comput. Methods Appl. Mech. Engrg. 196 (21-24) (2007) 2486-2496.

[15] L.P. Zhu, I. Elishakoff, J.H. Starnes Jr. Derivation of multi-dimensional ellipsoidal convex model for experimental data, Math. Comput. Modelling 24 (2) (1996) 103-114.

[16] C. Jiang, X. Han, G.Y. Lu, J. Liu, Z. Zhang, Y.C. Bai, Correlation analysis of non-probabilistic convex model and corresponding structural reliability technique, Comput. Methods Appl. Mech. Engrg. 200 (33-36) (2011) 2528–2546.

[17] Z. Kang, W.B. Zhang, Construction and application of an ellipsoidal convex model using a semi-definite programming formulation from measured data, Comput. Methods Appl. Mech. Engrg. 300 (2016) 461-489.

[18] C. Jiang, R.G. Bi, G.Y. Lu, X. Han, Structural reliability analysis using non-probabilistic convex model, Comput. Methods Appl. Mech. Engrg. 254 (2013) 83-98.

[19] Z.P. Qiu, X.J. Wang, Two non-probabilistic set-theoretical models for dynamic response and buckling failure measures of bars with unknown-but-bounded initial imperfections, Int. J. Solids Struct. 42 (3-4) (2005) 1039-1054.

[20] Y.J. Luo, Z. Kang, Z. Luo, A. Li, Continuum topology optimization with non-probabilistic reliability constraints based on multi-ellipsoid convex model, Struct. Multidisc. Optim. 39 (3) (2009) 297-310.

[21] Z. Kang, Y.J. Luo, Non-probabilistic reliability-based topology optimization of geometrically nonlinear structures using convex models, Comput. Methods Appl. Mech. Engrg. 198 (41-44) (2009) 3228-3238.

[22] C. Jiang, Q.F. Zhang, X. Han, Multidimensional parallelepiped model — a new type of non-probabilistic convex model for structural uncertainty quantification, Int. J. Numer. Meth. Eng. 103 (1) (2015) 31-59.

[23] B.Y. Ni, C. Jiang, X. Han, An improved multidimensional parallelepiped non-probabilistic model for structural uncertainty analysis. Appl. Math. Model. 40 (7-8) (2016) 4727-4745.

[24] I. Elishakoff, Y. Bekel, Application of Lamé's super ellipsoids to model initial imperfections, ASME J. Appl. Mech. 80 (6) (2013) 061006-1.

[25] C. Jiang, B.Y. Ni, X. Han, Y.R. Tao, Non-probabilistic convex model process: A new method of time-variant uncertainty analysis and its application to structural dynamic reliability problems, Comput. Methods Appl. Mech. Engrg. 268 (2014) 656-676.

[26] D. Moens, D. Vandepitte, A survey of non-probabilistic uncertainty treatment in finite element analysis, Comput. Methods Appl. Mech. Engrg. 194 (12-16) (2005) 1527-1555.

[27] R.L. Muhanna, H. Zhang, R.L. Mullen, Interval finite elements as a basis for generalized models of uncertainty in engineering mechanics, Reliable Comput. 13 (2) (2007) 173-194.

[28] H.J. Cao, B.Y. Duan, An approach on the non-probabilistic reliability of structures based on uncertainty convex models, Chin. J. Comput. Mech. 22 (5) (2005) 546-549.

[29] C. Jiang, Q.F. Zhang, X. Han, Y.H. Qian, A non-probabilistic structural reliability analysis method based on a multidimensional parallelepiped convex model, Acta. Mech. 225 (2) (2014) 383-395.

[30] L. Wang, X.J. Wang, R.X. Wang, X. Chen, Time-dependent reliability modeling and analysis method for mechanics based on convex process, Math. Probl. Eng. 2015, (2015) 1-16.

[31] X.J. Wang, L. Wang, I. Elishakoff, Z.P. Qiu, Probability and convexity concepts are not

antagonistic, Acta. Mech. 219 (1-2) (2011) 45-64.

[32] C. Jiang, Z.G. Zhang, Q.F. Zhang, X. Han, H.C. Xie, J. Liu, A new nonlinear interval programming method for uncertain problems with dependent interval variables, Eur. J. Oper. Res. 238 (1) (2014) 245-253.

[33] X.P. Du, Reliability-based design optimization with dependent interval variables, Int. J. Numer. Meth. Eng. 91 (2) (2012) 218-228.

[34] S. Ganzerli, C.P. Pantelides, Optimum structural design via convex model superposition, Comput. Struct. 74 (1) (2000) 639-647.

[35] M. Lombardi, Optimization of uncertain structures using non-probabilistic models, Comput. Struct. 67 (1-3) (1998) 99-103.

[36] J.L. Wu, Z. Luo, N. Zhang, Y.Q. Zhang, A new interval uncertain optimization method for structures using Chebyshev surrogate models, Comput. Struct. 146 (2015) 185-196.

[37] L. Wang, X.J. Wang, R.X. Wang, X. Chen, Reliability-based design optimization under mixture of random, interval and convex uncertainties, Arch. Appl. Mech. 86 (7) (2016) 1341-1367.

[38] X.P. Du, Interval reliability analysis, in: ASME 2007 Design Engineering Technical Conference & Computers and Information in Engineering Conference (DETC2007), Las Vegas, Nevada, USA, 2007.

[39] W. Gao, C.M. Song, F. Tin-Loi, Probabilistic interval response and reliability analysis of structures with a mixture of random and interval properties, Comput. Model. Eng. Sci. 46 (2) (2009) 151-189.

[40] F. Tonon, A. Bernardini, I. Elishakoff, Hybrid analysis of uncertainty: probability, fuzziness and anti-optimization, Chaos Solitons Fract. 12 (8) (2001) 1403-1414.

[41] W. Gao, D. Wu, C.M. Song, F. Tin-Loi, X.J. Li, Hybrid probabilistic interval analysis of bar structures with uncertainty using a mixed perturbation Monte-Carlo method, Finite Elem. Anal. Des. 47 (2011) 643-652.

[42] N.J. Higham, Computing real square roots of a real matrix, Linear Algebra Appl. 88-89 (1987) 405-430.

[43] J. Michael Steele, The Cauchy-Schwarz Master Class: An Introduction to the Art of Mathematical Inequalities, Cambridge University Press, Cambridge, 2004.

[44] X.T. Feng, Introduction to Intelligent Rock Mechanics, Science Press, Beijing, 2000.

[45] W. Barfield, Fundamentals of wearable computers and augmented reality, second ed., CRC Press, Boca Raton, 2015.

[46] Z.L. Huang, C. Jiang, Z. Zhang, T. Fang, X. Han, A decoupling approach for evidence-theory-based reliability design optimization, Struct. Multidisc. Optim. 56 (2017) 647-661.

# Figure captions

# Figure captions

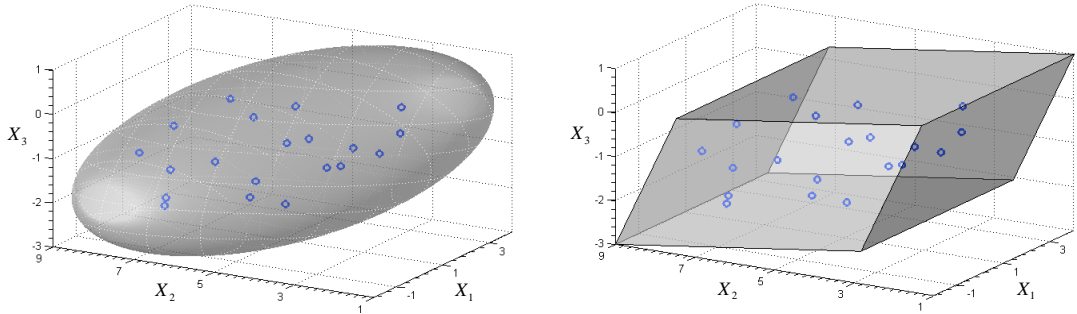

(a) ME model       (b) MP model

**Fig. 1.** ME model and MP model.

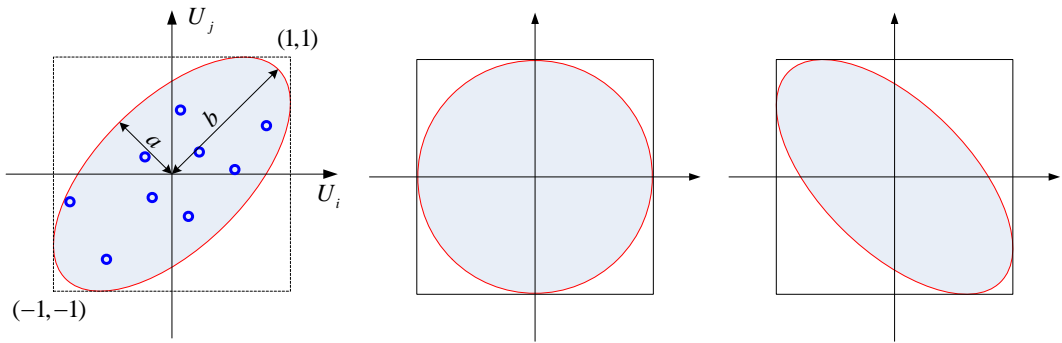

(a) Positively correlated    (b) Uncorrelated    (c) Negatively correlated

**Fig. 2.** Correlation quantification in ME model.

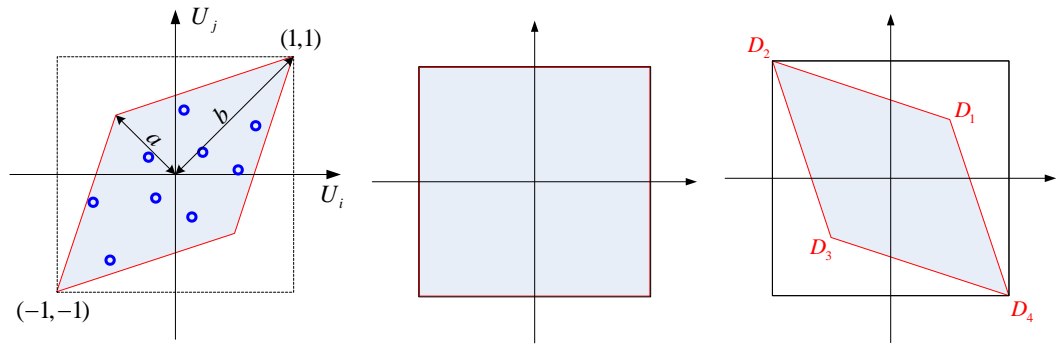

(a) Positively correlated    (b) Uncorrelated    (c) Negatively correlated

**Fig. 3.** Correlation quantification in MP-II model.

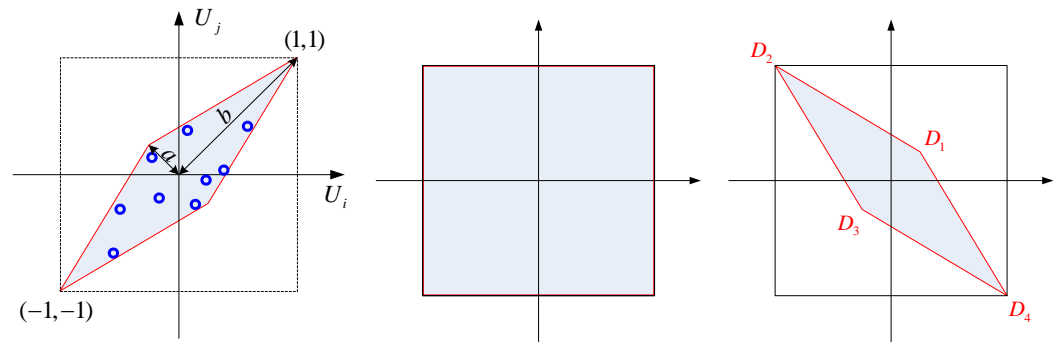

(a) Positively correlated    (b) Uncorrelated    (c) Negatively correlated

**Fig. 4.** Correlation quantification in MP-I model.

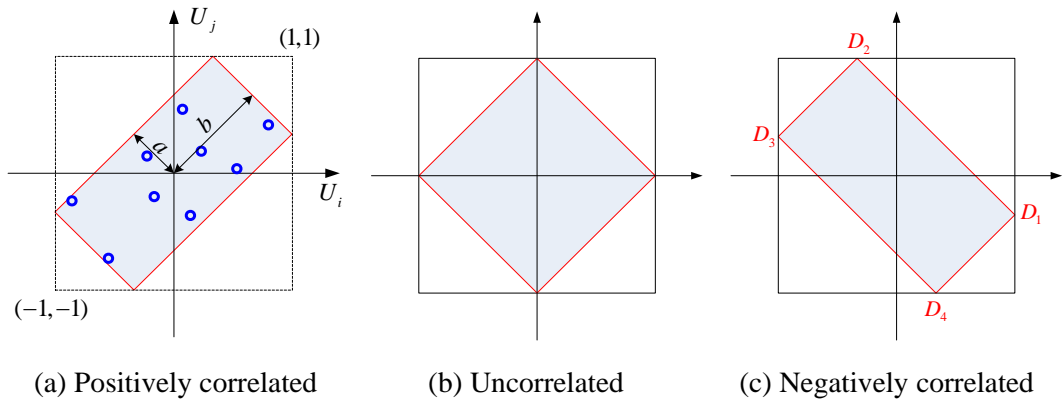

(a) Positively correlated  (b) Uncorrelated  (c) Negatively correlated

**Fig. 5.** Correlation quantification in Rectangular MP model.

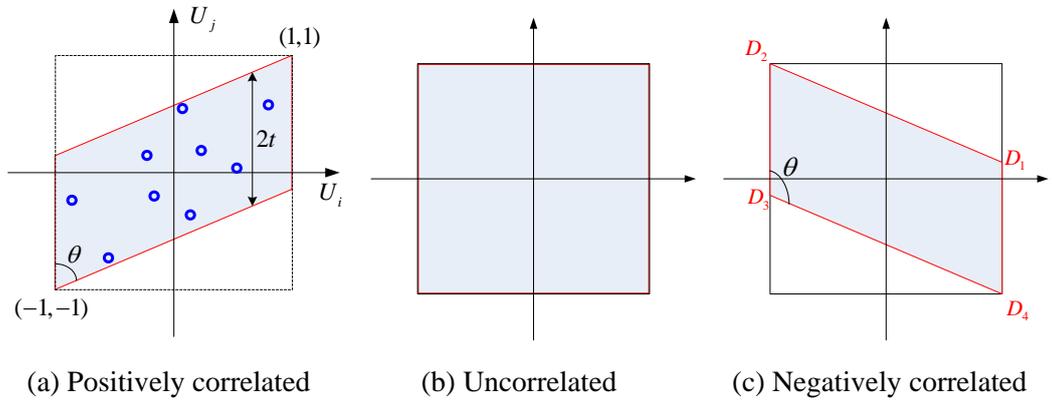

(a) Positively correlated  (b) Uncorrelated  (c) Negatively correlated

**Fig. 6.** Correlation quantification in LTri-MP model.

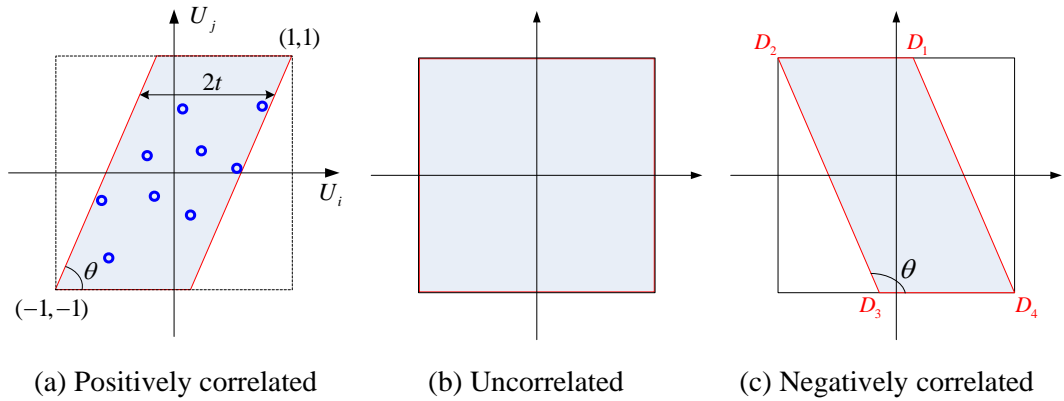

(a) Positively correlated  (b) Uncorrelated  (c) Negatively correlated

**Fig. 7.** Correlation quantification in UTri-MP model.

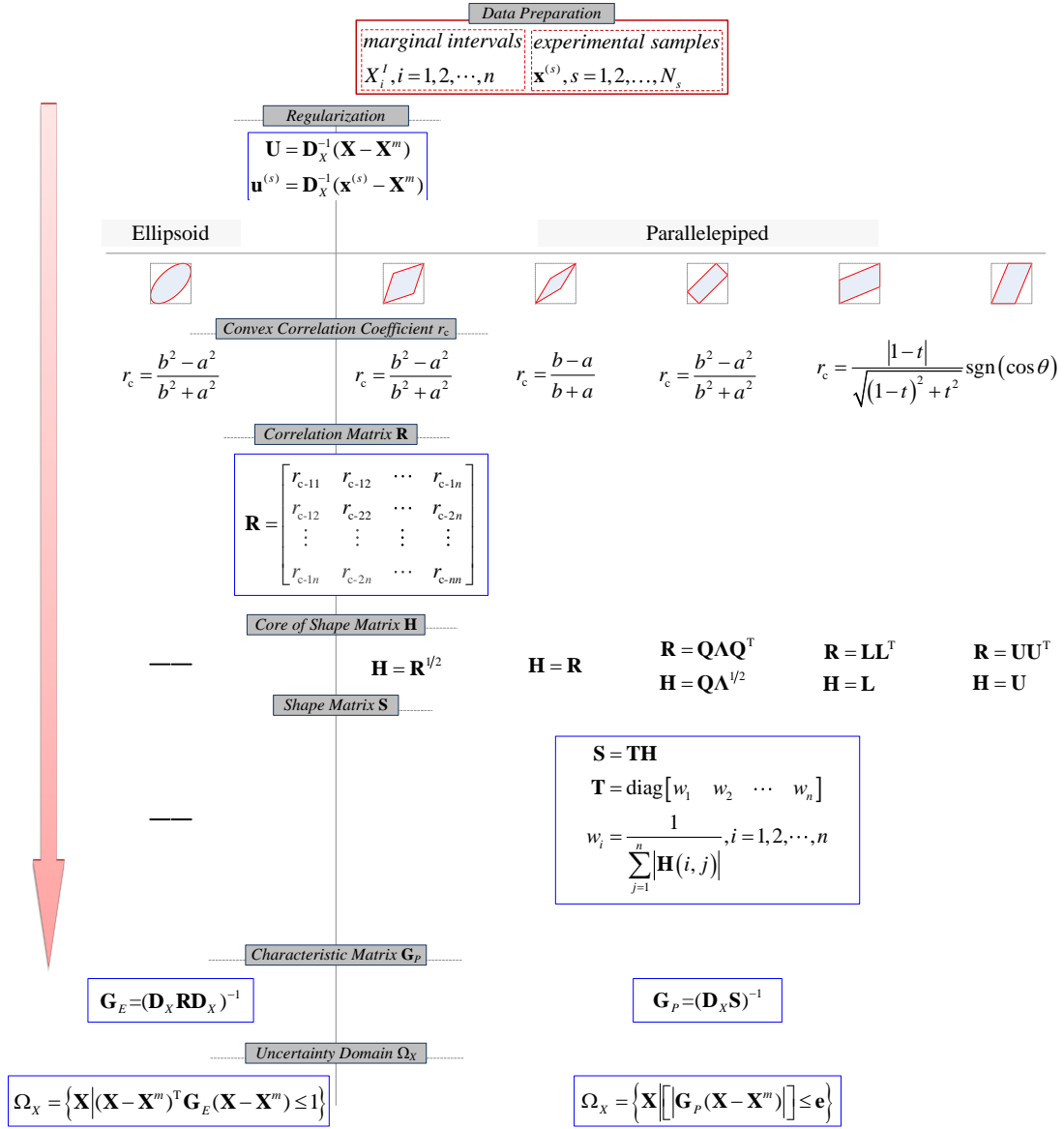

**Fig. 8.** A unified construction procedure for the convex models.

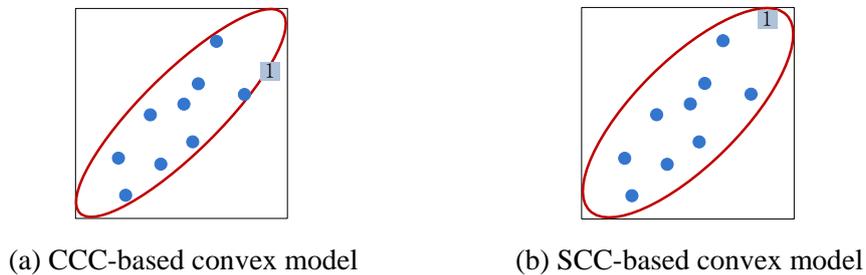

(a) CCC-based convex model      (b) SCC-based convex model

**Fig. 9.** The first difference between SCC and CCC in convex modelling.

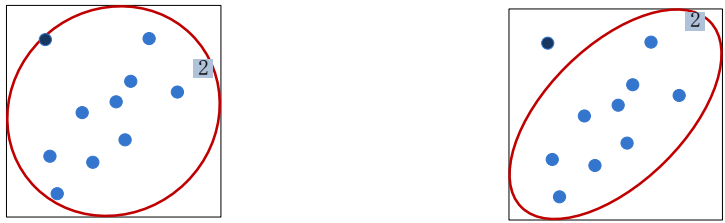

(a) CCC-based convex model      (b) SCC-based convex model

**Fig. 10.** The second difference between SCC and CCC in convex modelling.

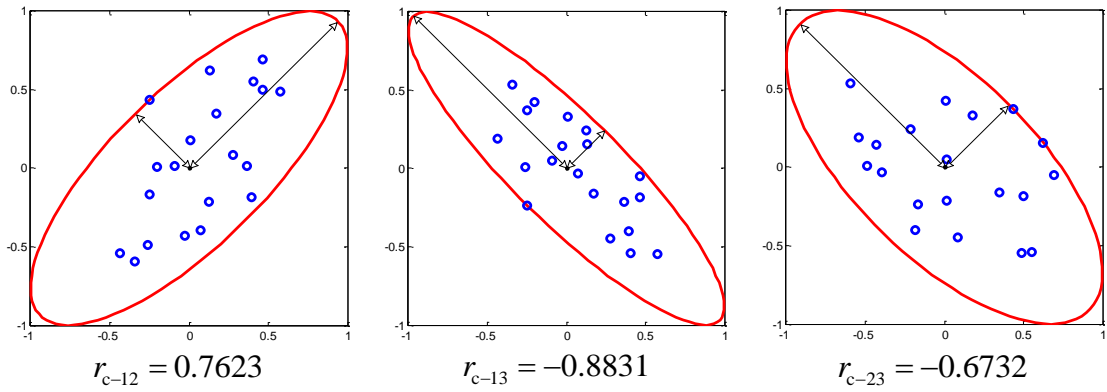

$r_{c-12} = 0.7623$      $r_{c-13} = -0.8831$      $r_{c-23} = -0.6732$

**Fig. 11.** CCCs of $U_1$ and $U_2$, $U_1$ and $U_3$, $U_2$ and $U_3$ in ME model.

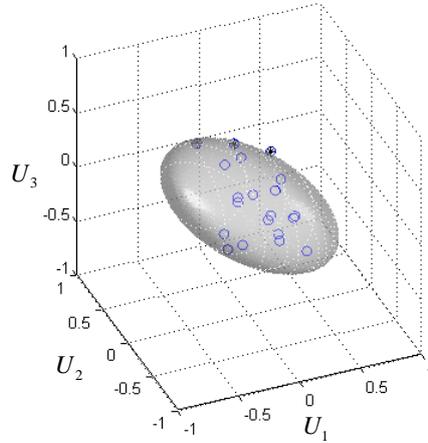

(a) 3D Ellipsoid-shaped uncertainty domain

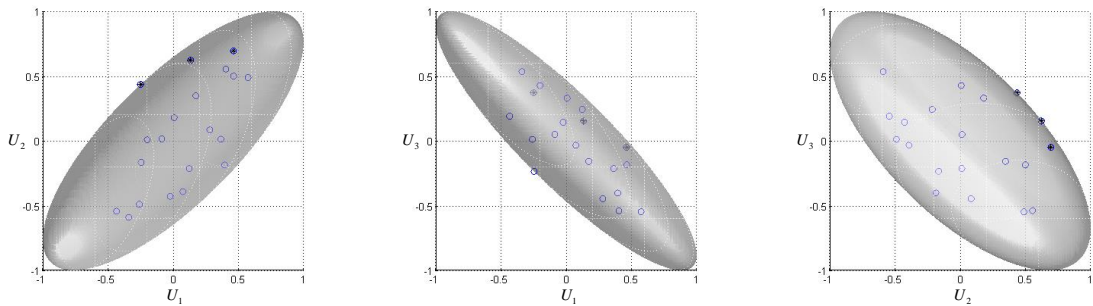

(b) Projection on $U_1$-$U_2$ plane (c) Projection on $U_1$-$U_3$ plane (d) Projection on $U_2$-$U_3$ plane

**Fig. 12.** 3D Ellipsoid-shaped uncertainty domain and its projections (CCC).

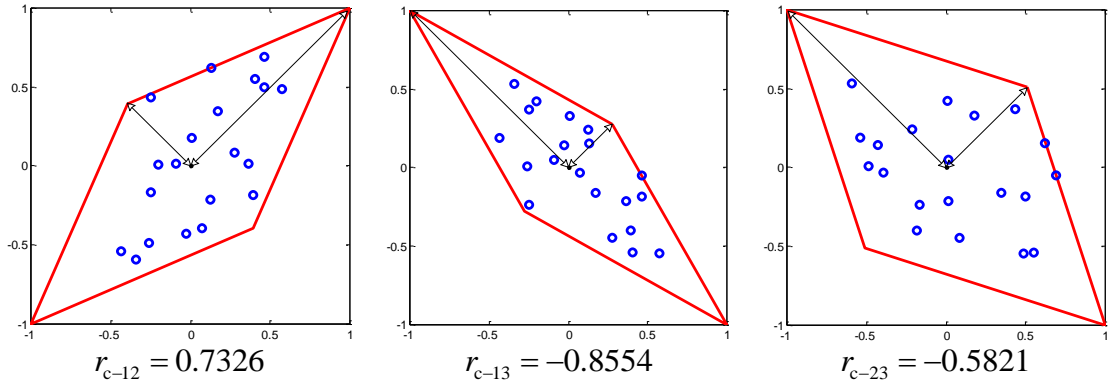

**Fig. 13.** CCCs of $U_1$ and $U_2$, $U_1$ and $U_3$, $U_2$ and $U_3$ in MP-II model.

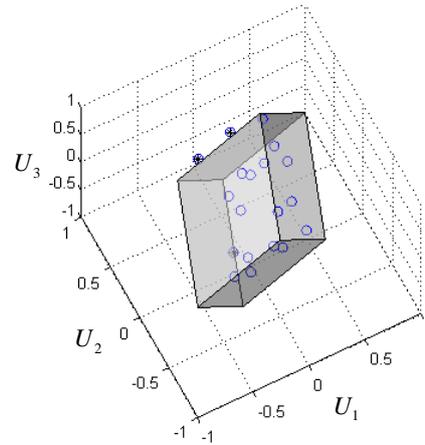

(a) 3D MP-II-shaped uncertainty domain

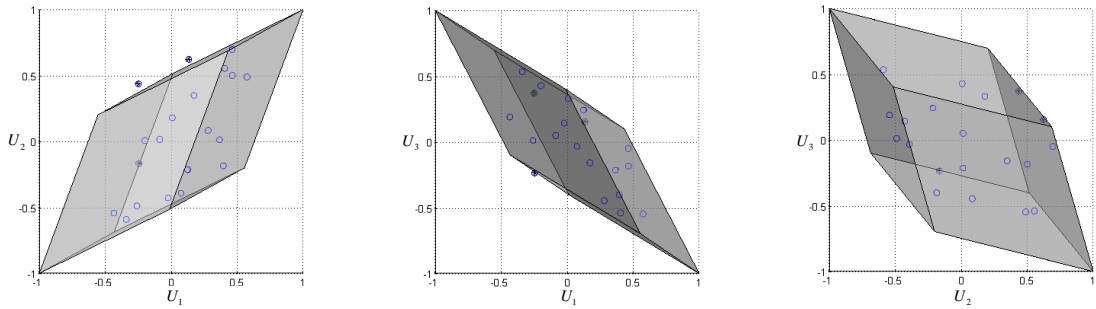

(b) Projection on $U_1$-$U_2$ plane (c) Projection on $U_1$-$U_3$ plane (d) Projection on $U_2$-$U_3$ plane
**Fig. 14.** 3D MP-II-shaped uncertainty domain and its projections (CCC).

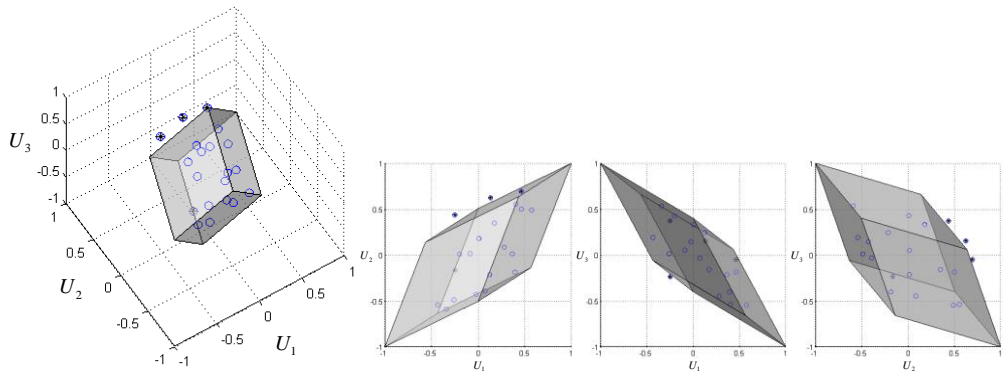

(a) 3D MP-I-shaped uncertainty domain and its projections (CCC)

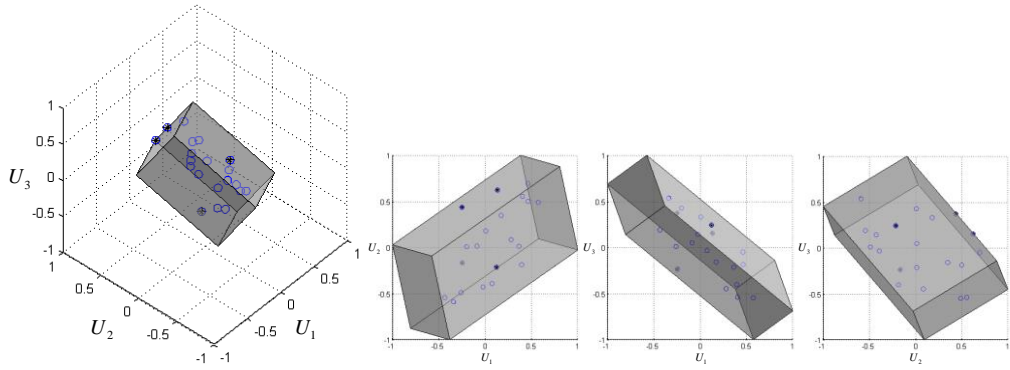

(b) 3D Rectangular MP-shaped uncertainty domain and its projections (CCC)

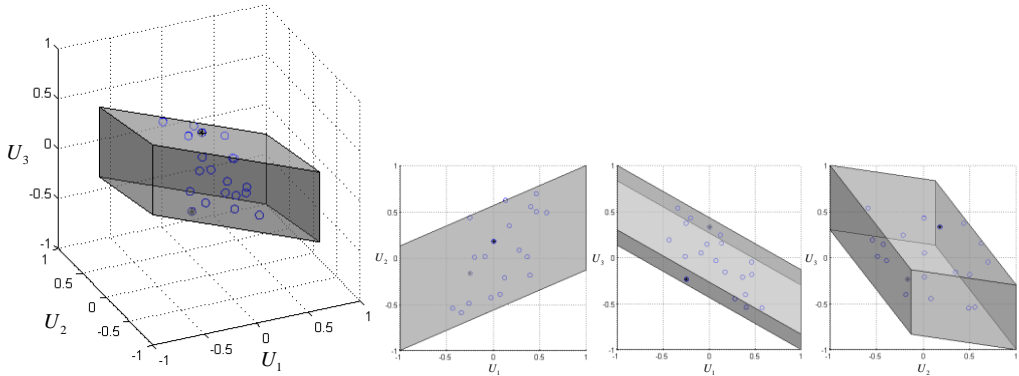

(c) 3D LTri-MP-shaped uncertainty domain and its projections (CCC)

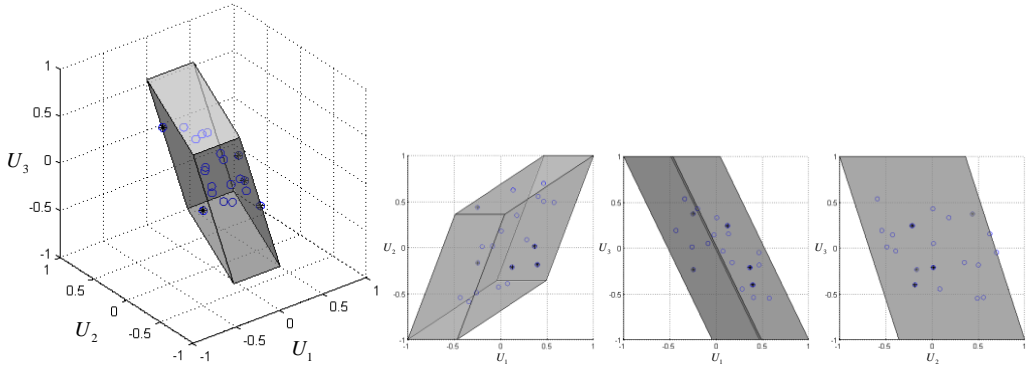

(d) 3D UTri-MP-shaped uncertainty domain and its projections (CCC)

**Fig. 15.** 3D uncertainty domains and their projections depicted by the other MP models (CCC).

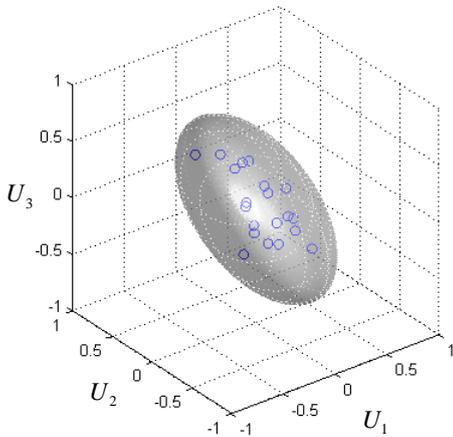

(a) ME-shaped uncertainty domain

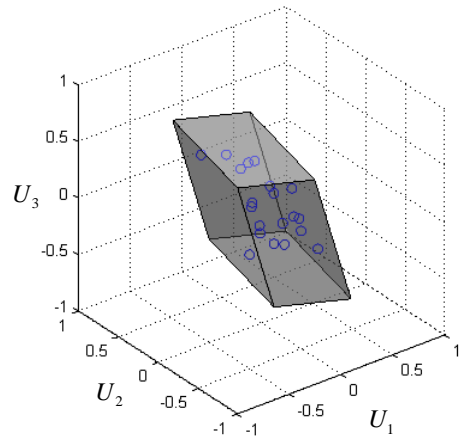

(b) MP-II-shaped uncertainty domain

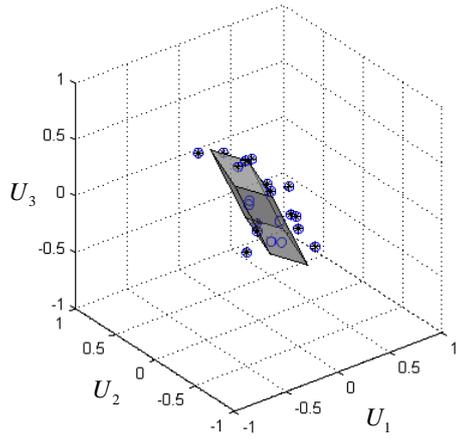
(c) MP-I-shaped uncertainty domain

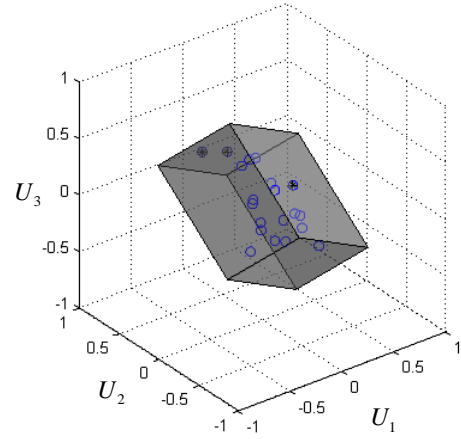
(d) Rectangular MP-shaped uncertainty domain

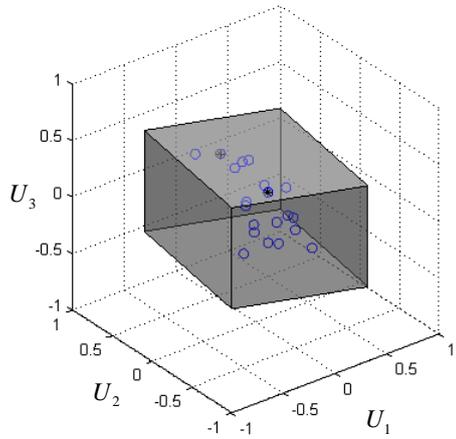
(e) LTri-MP-shaped uncertainty domain

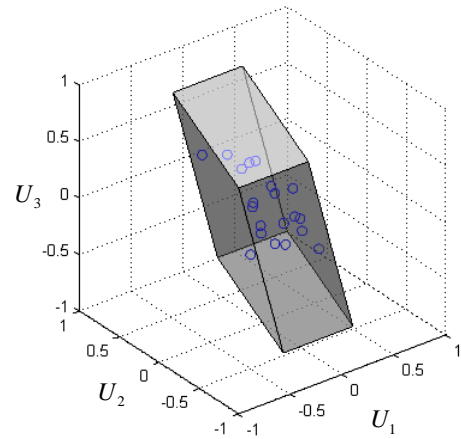
(f) UTri-MP-shaped uncertainty domain

**Fig. 16.** 3D Uncertainty domains depicted by the convex models (SCC).

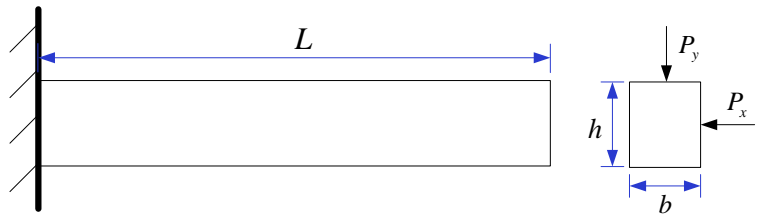

**Fig. 17.** A cantilever beam [18].

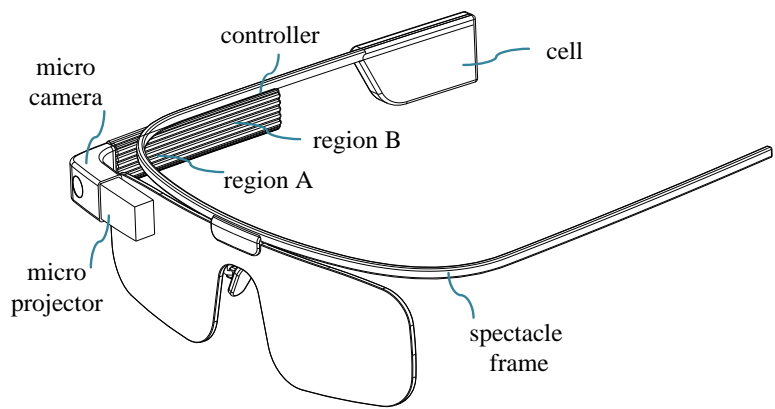

(a) Components of the augmented reality glasses

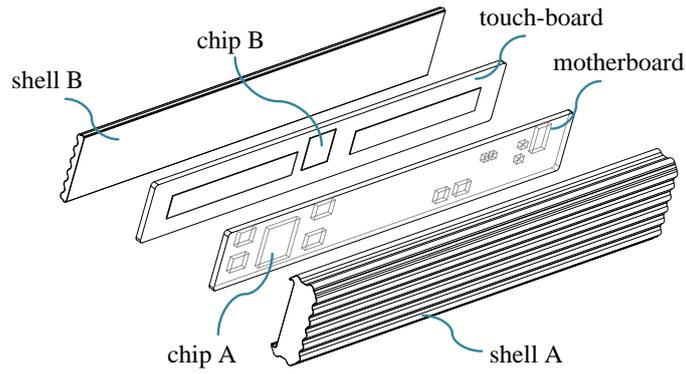

(b) Exploded diagram of the controller

**Fig. 18.** An augmented reality glasses [46].

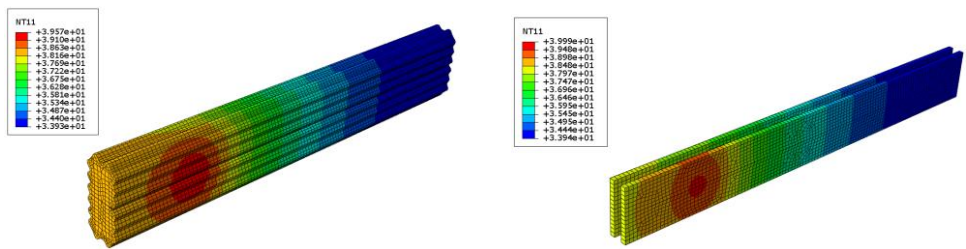

(a) Surface temperature of the shell      (b) Temperature of the circuit board

**Fig. 19.** Numerical simulation model of the controller of the AR glasses [46].

**Table 1** Unbiasedness property of the convex modelling methods.

| Shape | Convex Models | Mathematical Expression | Convex Correlation Coefficient | Sample Correlation Coefficient | UnBiasedness (CCC) | UnBiasedness (SCC) |
|---|---|---|---|---|---|---|
| 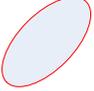 | ME model | $(\mathbf{X}-\mathbf{X}^m)^\mathrm{T}\mathbf{G}_E(\mathbf{X}-\mathbf{X}^m) \leq 1$ | $r_c = \dfrac{b^2-a^2}{b^2+a^2}$ | $r_s = \dfrac{\sum_{s=1}^{N_s}\left(x_i^{(s)}-X_i^m\right)\left(x_j^{(s)}-X_j^m\right)}{\sqrt{\sum_{s=1}^{N_s}\left(x_i^{(s)}-X_i^m\right)^2 \sum_{s=1}^{N_s}\left(x_j^{(s)}-X_j^m\right)^2}}$ | UnBiased | UnBiased |
| 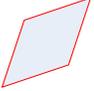 | MP-II model | | $r_c = \dfrac{b^2-a^2}{b^2+a^2}$ | | Biased | UnBiased |
| 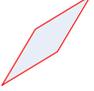 | MP-I model | | $r_c = \dfrac{b-a}{b+a}$ | | Biased | Biased |
| 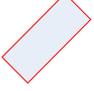 | Rectangular MP model | $\left[\left|\mathbf{G}_P(\mathbf{X}-\mathbf{X}^m)\right|\right] \leq \mathbf{e}$ | $r_c = \dfrac{b^2-a^2}{b^2+a^2}$ | | Biased | UnBiased |
| 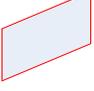 | LTri-MP model | | $r_c = \dfrac{|1-t|}{\sqrt{(1-t)^2+t^2}}\mathrm{sgn}(\cos\theta)$ | | Biased | UnBiased |
| 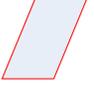 | UTri-MP model | | $r_c = \dfrac{|1-t|}{\sqrt{(1-t)^2+t^2}}\mathrm{sgn}(\cos\theta)$ | | Biased | UnBiased |

**Table 2** Hypothetical samples of $U_1$, $U_2$ and $U_3$.

| No. | $U_1$ | $U_2$ | $U_3$ | No. | $U_1$ | $U_2$ | $U_3$ |
|---|---|---|---|---|---|---|---|
| 1 | 0.365 | 0.012 | -0.213 | 11 | 0.074 | -0.392 | -0.034 |
| 2 | -0.340 | -0.590 | 0.533 | 12 | 0.574 | 0.488 | -0.547 |
| 3 | 0.463 | 0.499 | -0.185 | 13 | -0.434 | -0.543 | 0.190 |
| 4 | 0.133 | 0.622 | 0.153 | 14 | -0.259 | -0.489 | 0.011 |
| 5 | 0.123 | -0.214 | 0.242 | 15 | 0.172 | 0.347 | -0.159 |
| 6 | -0.246 | -0.166 | -0.236 | 16 | 0.007 | 0.179 | 0.330 |
| 7 | -0.023 | -0.428 | 0.143 | 17 | 0.394 | -0.186 | -0.402 |
| 8 | -0.249 | 0.435 | 0.370 | 18 | 0.280 | 0.083 | -0.447 |
| 9 | 0.403 | 0.553 | -0.539 | 19 | 0.462 | 0.694 | -0.049 |
| 10 | -0.087 | 0.014 | 0.049 | 20 | -0.199 | 0.007 | 0.426 |

**Table 3** Evaluation indexes of the constructed convex models (CCC).

| Convex Model | $\kappa$ | $\nu$ | $\bar{\nu}$ |
|---|---|---|---|
| ME | 18/20 | 15.90% | 54.18% |
| MP-II | 17/20 | 9.17% | 45.10% |
| MP-I | 16/20 | 8.26% | 43.54% |
| Rectangular MP | 16/20 | 14.74% | 52.82% |
| LTri-MP | 18/20 | 19.79% | 58.28% |
| UTri-MP | 15/20 | 18.07% | 56.54% |

**Table 4** Evaluation indexes of the constructed convex models (SCC).

| Convex Model | $\kappa$ | $\nu$ | $\bar{\nu}$ |
|---|---|---|---|
| ME | 20/20 | 27.86% | 65.31% |
| MP-II | 20/20 | 17.33% | 55.75% |
| MP-I | 5/20 | 2.97% | 30.97% |
| Rectangular MP | 17/20 | 16.37% | 54.70% |
| LTri-MP | 18/20 | 24.51% | 62.58% |
| UTri-MP | 20/20 | 24.49% | 62.57% |

**Table 5** Experimental samples of the geotechnical parameters [44].

| No. | $X_1$ ($p$/%) | $X_2$ ($W_1$/%) | $X_3$ ($I_p$/%) | $X_4$ ($B$/m$^2$ g$^{-1}$) | $X_5$ ($C$/%) | $Y$ ($f_r$) |
|---|---|---|---|---|---|---|
| 1 | 53 | 31 | 12 | 75 | 26.99 | 0.45 |
| 2 | 47 | 23 | 8 | 74 | 81.44 | 0.51 |
| 3 | 52 | 44 | 22 | 138 | 31.26 | 0.37 |
| 4 | 60 | 34 | 16 | 202 | 9.06 | 0.4 |
| 5 | 34 | 20 | 6 | 31 | 23.03 | 0.52 |
| 6 | 49 | 30 | 15 | 65 | 16.2 | 0.38 |
| 7 | 38 | 18 | 9 | 58 | 15.7 | 0.45 |
| 8 | 35 | 26 | 10 | 34 | 35.5 | 0.54 |
| 9 | 43 | 27 | 11 | 55 | 12.14 | 0.53 |
| 10 | 51 | 28 | 10 | 36.39 | 36.39 | 0.51 |

**Table 6** Variation intervals of the geotechnical parameters.

| uncertain parameters | variation interval | midpoint | radius |
|---|---|---|---|
| clay content $X_1$ ($p$/%) | [28, 66] | 47 | 19 |
| liquid limit $X_2$ ($W_l$/%) | [12, 50] | 31 | 19 |
| plasticity index $X_3$ ($I_p$/%) | [2, 26] | 14 | 12 |
| specific surface area $X_4$ ($B$/m$^2$ g$^{-1}$) | [0, 240] | 120 | 120 |
| carbonate content $X_5$ ($C$/%) | [0, 100] | 50 | 50 |
| residual strength $Y$ ($f_r$) | [0.3, 0.6] | 0.45 | 0.15 |

**Table 7** Evaluation indexes of the constructed convex models for the geotechnical parameters.

| convex model | $\kappa$ | $\nu$ | $\bar{\nu}$ |
|---|---|---|---|
| ME model (CCC) | 10/10 | 0.53% | 41.70% |
| MP-II model (CCC) | 0/10 | 0.07% | 29.97% |
| ME model (SCC) | 3/10 | 0.32% | 38.42% |
| MP-II model (SCC) | 0/10 | 0.08% | 30.52% |

**Table 8** Samples of the geometrical parameters of the beam.

| No. | $b$ | $h$ | $L$ | No. | $b$ | $h$ | $L$ |
|---|---|---|---|---|---|---|---|
| 1 | 97.69 | 206.31 | 926.14 | 17 | 91.83 | 196.72 | 965.26 |
| 2 | 94.41 | 199.87 | 957.64 | 18 | 105.90 | 207.54 | 990.00 |
| 3 | 102.21 | 190.99 | 1021.50 | 19 | 94.24 | 187.77 | 961.88 |
| 4 | 102.93 | 210.95 | 951.07 | 20 | 106.43 | 200.40 | 979.21 |
| 5 | 93.51 | 193.42 | 963.15 | 21 | 103.22 | 186.30 | 959.35 |
| 6 | 106.17 | 203.32 | 1045.33 | 22 | 105.62 | 196.03 | 1067.40 |
| 7 | 100.86 | 202.23 | 1041.28 | 23 | 94.18 | 200.84 | 987.31 |
| 8 | 103.90 | 193.14 | 954.14 | 24 | 101.87 | 187.02 | 1037.59 |
| 9 | 95.97 | 208.08 | 974.42 | 25 | 98.08 | 187.99 | 978.36 |
| 10 | 103.81 | 207.95 | 1069.66 | 26 | 95.36 | 202.15 | 991.77 |
| 11 | 101.98 | 196.86 | 1019.12 | 27 | 98.54 | 198.42 | 1023.88 |
| 12 | 97.08 | 198.40 | 974.78 | 28 | 104.58 | 190.56 | 1023.37 |
| 13 | 101.32 | 202.56 | 956.40 | 29 | 95.62 | 205.75 | 1042.84 |
| 14 | 99.18 | 208.92 | 943.92 | 30 | 98.44 | 198.11 | 993.78 |
| 15 | 98.01 | 214.03 | 992.20 | 31 | 102.26 | 202.30 | 1056.27 |
| 16 | 94.87 | 191.42 | 1024.99 | 32 | 103.18 | 195.30 | 931.20 |

**Table 9** Variation intervals of the uncertain parameters of the AR glasses.

| uncertain parameters | variation interval | midpoint | radius |
|---|---|---|---|
| environmental temperature $T_a$ | [-10, 50] | 20 | 30 |
| air velocity $V_a$ | [0, 0.7] | 0.35 | 0.35 |
| power dissipation of chip A $P_A$ | [0, 0.6] | 0.3 | 0.3 |
| power dissipation of chip B $P_B$ | [0, 0.2] | 0.1 | 0.1 |